\documentclass[%
10pt,
superscriptaddress,
amsmath,amssymb,
aps,
prl,
twocolumn,
longbibliography
]{revtex4-2}

\usepackage{graphicx}
\usepackage{dcolumn}
\usepackage{bm}
\usepackage[colorlinks,urlcolor=blue,citecolor=blue,linkcolor=blue]{hyperref}

\usepackage{abraces}
\usepackage{braket}
\usepackage{cancel}
\usepackage[]{diffcoeff}
\usepackage{enumitem}
\usepackage{mathtools}
\usepackage{esint} 
\usepackage{interval}
\usepackage{lipsum}
\usepackage[version=4]{mhchem} 
\usepackage{nicematrix}
\usepackage[super]{nth} 
\usepackage{rotating}
\usepackage[separate-uncertainty=true,per-mode=symbol,multi-part-units=single,round-pad=false]{siunitx}
\usepackage{tensor} 
\usepackage{upgreek}
\usepackage{xfrac} 

\usepackage{lmodern}
\usepackage{anyfontsize}

\newcommand{\mycomment}[1]{}

\DeclarePairedDelimiter{\abs}{\lvert}{\rvert}

\DeclarePairedDelimiterX{\commutator}[2]{[}{]}{#1, #2}
\DeclarePairedDelimiterX{\anticomm}[2]{\{}{\}}{#1, #2}

\DeclareSIUnit\rydberg{Ry}

\difdef{f, s, c, l} {}
{
	op-symbol = d,
	op-order-nudge = 1 mu
}

\difdef{l}{delta}
{
	op-symbol = \delta
}

\difdef{l}{Delta}
{
	op-symbol = \Delta
}

\difdef{l}{D}
{
	op-symbol = \mathrm{D}
}

\intervalconfig{
	soft open fences ,
}

\intervalconfig{
	soft open fences ,
}

\graphicspath{{images/}}

\newcommand{\euler}{\ensuremath{e}}

\newcommand{\kB}{\ensuremath{k_\mathrm{B}}}
\newcommand{\muB}{\ensuremath{\mu_\mathrm{B}}}

\newcommand{\Ham}{\ensuremath{\mathcal{H}}}

\newcommand{\Zfun}{\ensuremath{\mathcal{Z}}}

\newcommand{\Bper}{\ensuremath{B_\perp}}

\newcommand{\Bcross}{\ensuremath{B^\times}}

\newcommand{\DSO}{\ensuremath{\Delta_\mathrm{SO}}}

\newcommand{\Kdown}{\ensuremath{\ket{K^-\downarrow}}}
\newcommand{\Kup}{\ensuremath{\ket{K^+\uparrow}}}
\newcommand{\GStwo}{\ensuremath{\ket{S_v T_s^0} \oplus \ket{T_v^0 S_s}}}

\newcommand{\Vpg}{\ensuremath{V_\mathrm{PG}}}
\newcommand{\VL}{\ensuremath{V_\mathrm{LB}}}
\newcommand{\VR}{\ensuremath{V_\mathrm{RB}}}

\newcommand{\VSD}{\ensuremath{V_\mathrm{SD}}}

\newcommand{\Idet}{\ensuremath{I_\mathrm{det}}}
\newcommand{\Ioff}{\ensuremath{I_\mathrm{off}}}
\newcommand{\Iheater}{\ensuremath{I_\mathrm{heater}}}

\newcommand{\Tbar}{\ensuremath{\overline{T}}}
\newcommand{\Nbar}{\ensuremath{\overline{N}}}

\newcommand{\apg}{\ensuremath{\alpha_\mathrm{PG}}}
\newcommand{\gm}{\ensuremath{g_\mathrm{m}}}
\newcommand{\GammaR}{\ensuremath{\Gamma}}

\begin{document}

	\preprint{APS/123-QED}

	\title{Entropy spectroscopy of a bilayer graphene quantum dot} 

	\author{C. Adam}
	\email{adamc@phys.ethz.ch}
	\author{H. Duprez}
	\author{N. Lehmann}
	\author{A. Yglesias}
	\author{A.~O.~Denisov}
	\author{S. Cances}
	\author{M. J. Ruckriegel}
	\author{M. Masseroni}
	\author{C. Tong}
	\author{W. Huang}
	\author{D. Kealhofer}
	\author{R. Garreis}
	\affiliation{Solid State Physics Laboratory, ETH Zurich, Zurich CH-8093, Switzerland}
	\author{K. Watanabe}
	\affiliation{Research Center for Electronic and Optical Materials, National Institute for Materials Science, 1-1 Namiki, Tsukuba 305-0044, Japan}
	\author{T. Taniguchi}
	\affiliation{Research Center for Materials Nanoarchitectonics, National Institute for Materials Science,  1-1 Namiki, Tsukuba 305-0044, Japan}
	\author{K. Ensslin}
	\author{T. Ihn}
	\affiliation{Solid State Physics Laboratory, ETH Zurich, Zurich CH-8093, Switzerland}

	\date{\today}

	\begin{abstract}	
	We measure the entropy change of charge transitions in an electrostatically defined quantum dot in bilayer graphene. Entropy provides insights into the equilibrium thermodynamic properties of both ground and excited states beyond transport measurements. For the one-carrier regime, the obtained entropy shows that the ground state has a two-fold degeneracy lifted by an out-of-plane magnetic field. This observation is in agreement with previous direct transport measurements and confirms the applicability of this novel method. For the two-carrier regime, the extracted entropy indicates a non-degenerate ground state at zero magnetic field, contrary to previous studies suggesting a three-fold degeneracy. We attribute the degeneracy lifting to the effect of Kane-Mele type spin--orbit interaction on the two-carrier ground state, which has not been observed before. Our work demonstrates the validity and efficacy of entropy measurements as a unique, supplementary experimental tool to investigate the degeneracy of the ground state in quantum devices build in materials such as graphene. This technique, applied to exotic systems with fractional ground state entropies, will be a powerful tool in the study of quantum matter.
\end{abstract}


	\maketitle

	\section{\label{sec:intro} Introduction}
Entropy is a cornerstone concept in physics offering profound insights into thermodynamic systems through its statistical interpretation and ties to uncertainty and information theory. Particularly fascinating – and seemingly paradoxical – is the use of entropy measurements in quantum systems to gain insights into our lack of knowledge about their microscopic quantum mechanical ground states such as the non-abelian fractional quantum Hall state at filling factor $5/2$~\cite{cooperObservableBulkSignatures2009,ben-shachDetectingNonAbelianAnyons2013}, Majorana fermions~\cite{smirnovMajoranaTunnelingEntropy2015} and multi-channel Kondo systems~\cite{hanFractionalEntropyMultichannel2022}. This distinctive property makes entropy a powerful tool for directly quantifying the unknown degeneracy of such ground states and for uncovering previously hidden properties within them.

In this Letter, we present an entropy measurement that leverages these properties to reveal a previously unnoticed lifting of the three-fold ground state degeneracy in a simple, confined, interacting two-particle system within bilayer graphene. Previous measurements of one-particle ground and excited states in gate-defined bilayer graphene quantum dots~\cite{kurzmannExcitedStatesBilayer2019, kurzmannKondoEffectSpin2021, mollerProbingTwoElectronMultiplets2021} consistently identified two spin-valley Kramers doublets separated by the small Kane–Mele spin–orbit coupling~\cite{kaneQuantumSpinHall2005, konschuhTheorySpinorbitCoupling2012} of $50 - 80 \unit{\micro\electronvolt}$. For the two-particle case, the ground state was previously determined to be a three-fold degenerate valley-singlet spin-triplet state~\cite{kurzmannExcitedStatesBilayer2019, mollerProbingTwoElectronMultiplets2021}, indicating that the Coulomb exchange interactions dominate over the spin–orbit splitting.
Here, using the low-temperature entropy measurement technique pioneered in Refs.~\cite{hartmanDirectEntropyMeasurement2018,childEntropyMeasurementStrongly2022,childRobustProtocolEntropy2022}, we demonstrate that this degenerate ground state undergoes further splitting resulting in a non-degenerate ground state. Theoretical analysis indicates that spin-orbit interactions split the spin-triplet state into $1+2$ sublevels, in agreement with our experimental findings. These results highlight the power of entropy measurements in quantum dot systems as an alternative to excited state spectroscopy and shed new light on the influence of spin–orbit coupling in two-particle graphene quantum dots.

	\section{\label{sec:device} Device \& Experimental Setup}
	\begin{figure}[t!]
		\includegraphics{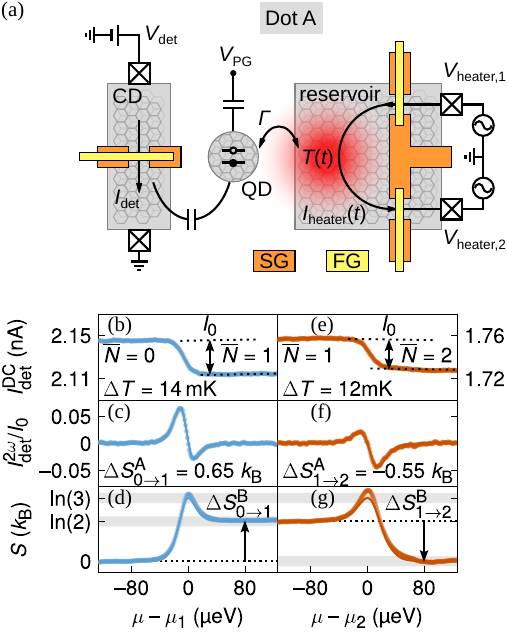}
		\caption{\label{fig:entropy}
			(a) The system consists of a quantum dot (QD) thermally coupled to a reservoir.
			The reservoir temperature $T(t)$ is changed by driving a current $\Iheater$ through an ohmic region in the BLG. This region is induced by the heater structure, consisting split gates (SG) and finger gates (FG). The charge detector (CD), capacitively coupled to the QD, carries a current $\Idet$ which changes as the number of charge carriers changes on the QD.
			For the $0\to 1$- and $1\to 2$-transition, respectively: The DC component of the detector current (b) and (e) with the extracted temperature modulation (continuous traces are fits to eq.~\ref{eq:detDC}, the electronic background temperatures are $\Tbar_{0\to 1} = \SI{60}{\milli\kelvin}$ and $\Tbar_{1\to 2} = \SI{95}{\milli\kelvin}$), second harmonic current normalized to $I_0$ (c) and (f) with extracted entropy change (continuous traces are fits to eq.~\ref{eq:dethamtwo}), and the entropy obtained by integration (d) and (g) (continuous traces are plots of eq.~\ref{eq:S01} in (d) and eq.~\ref{eq:S12}\cite{supplemental2025} in (g)).
			The gray bars $\pm 0.1 \ln(2)$ as a guide to the eye.
		}

	\end{figure}

	The QD investigated in the main text is referred to as ``Dot A''. Measurements on a second QD (``Dot B'') are shown in the supplemental material~\cite{supplemental2025}. The system is shown in Fig.~\ref{fig:entropy}(a). It consists of a gate-defined QD in thermal equilibrium with a bath of hole carriers (reservoir), and a charge detection circuit (detector).
	The number of hole carriers on the QD is controlled with the plunger gate voltage $\Vpg$ by changing the addition energy $\mu_N$ required to add the $N$th carrier in its ground state, according to $\Delta\mu_{N}=\apg \abs{e} \Delta\Vpg$. The plunger gate leverarm $\apg \approx 0.025$ is extracted from finite bias spectroscopy measurements and found to be independent of the carrier number (see Figs.~\ref{fig:onehole}(a),~\ref{fig:twohole}(a)~\cite{supplemental2025}).
	A current $\Iheater$ is driven through electrostatically defined constrictions in the BLG called heaters. The ohmic resistance of the heaters causes power to be dissipated as heat, inducing a temperature change due to Joule heating~\cite{huardElectronHeatingMetallic2007}.
	A detailed description of the sample and its operation is shown in Figs.~\ref{fig:device},~\ref{fig:operation}~\cite{supplemental2025}, respectively. In the following, we ellaborate how the entropy of the QD can be extracted from the charge detection signal.

	\section{\label{sec:entropy} Entropy Extraction}
	
	The system is described by the mean occupation number $\Nbar$ and the entropy $S$ of the QD, and the electrochemical potential $\mu$ and mean carrier temperature $\Tbar$ of the reservoir. In our experiment, $\mu$ and $\Tbar$ are independent, externally controlled variables. At constant $\Tbar$, these variables are connected through the Maxwell relation
	\begin{equation}
		\label{eq:maxwell}
		S(\mu,\Tbar) - S(\mu_0,\Tbar) = \int^{\mu}_{\mu_0} \diffp{\Nbar}{T}[\mu,T=\Tbar] \dl{\mu}.
	\end{equation}
	By measuring $\difsp{\Nbar}{T}[\mu]$ one can get the entropy of the system. In order to measure this quantity, we control the temperature $T(t)$ by driving an electrical current
	\begin{equation}
		\Iheater(t) = \Iheater \cdot \cos(\omega t)
	\end{equation}
	through the heaters. The temperature resulting from Joule heating can be expressed as~\cite{huardElectronHeatingMetallic2007}
	\begin{align}
		T(t)
		\label{eq:Thyp}
		&= \sqrt{(T^\mathrm{base})^2 + a \Iheater^2(t)} \\
		\label{eq:Tpar}
		&\approx \Tbar + \dl.Delta.{T} \cos(2\omega t),
	\end{align}
	where $T^\mathrm{base}$ is the reservoir temperature without any heating current and $a$ is a system-specific constant which is determined by a calibration measurement (see Fig.~\ref{fig:heating} (a)).
	The average temperature increases to $\Tbar = T^\mathrm{base} + \dl.Delta.{T}$, while a temperature modulation $\dl.Delta.{T}$ with frequency $2\omega$ is introduced.  The frequency $\omega = \SI{40}{\hertz}$ is chosen sufficiently low to ensure that the QD is in equilibrium with the reservoir at all times. The charge detector (CD) capacitively coupled to the QD~\cite{kurzmannChargeDetectionGateDefined2019} allows us to extract the mean occupation number $\Nbar$ via the detector current $\Idet(t) = \Idet^\mathrm{DC} + \Idet^{2\omega}$ where
	\begin{align}
		\label{eq:detDC}
		\Idet^\mathrm{DC} &= I_0 \cdot \Nbar(\mu, \Tbar) + \Ioff + \gm \Vpg \\
		\label{eq:dethamtwo}
		\Idet^{2\omega} &= I_0 \cdot \diffp{\Nbar}{T}[\mu,T=\Tbar] \dl.Delta.{T} \cdot \cos(2\omega t)
	\end{align}
	where $I_0$ is the change in CD current in response to a charge change on the QD, $\Ioff$ the offset current through the CD at the operation point and $\gm$ the transconductance of the CD with respect to $\Vpg$.
	The entropy difference $\dl.Delta.{S}_{N-1\to N}$ between the integer charge states $N-1$ and $N$ is obtained in two different ways.
	
	\textbf{Method A:}
	This method is only valid in the thermally broadened regime, i.e. $\kB T(t) \gg \GammaR$. One can find the analytical expressions
	\begin{align}
		\label{eq:NT}
		\Nbar(\mu,\Tbar)
		&= \frac{1}{1 + e^{\frac{\mu_{N}-\mu-\Tbar\dl.Delta.{S}_{N-1 \to N}}{\kB \Tbar}}} + N - 1, \\
		\label{eq:dNdT}
		\diffp{\Nbar}{T}[\mu,T=\Tbar]
		&= \frac{\frac{\mu_{N}-\mu-\dl.Delta.{S}_{N-1 \to N}}{4 \kB \Tbar^2}}{\cosh^2\left(\frac{\mu_{N}- \mu -\Tbar\dl.Delta.{S}_{N-1 \to N}}{2 \kB \Tbar}\right)}.
	\end{align}
	which contains $\dl.Delta.{S}_{N-1\to N}$ as a parameter. Since $\Idet^{2\omega} \propto \difsp{\Nbar}{T}[\mu,T=\Tbar]$ one can obtain $\dl.Delta.{S}_{N-1\to N}$ by fitting eq.~\ref{eq:dNdT} to the curve shape of $\Idet^{2\omega}$. This can be done without performing a temperature calibration.
	
	\textbf{Method B:}
	This more general method requires the temperature modulation $\dl.Delta.{T}$ to be known. Then  eq.~\ref{eq:dethamtwo} allows to get $\difsp{\Nbar}{T}[\mu,T=\Tbar]$ directly from $\Idet^{2\omega}$. After obtaining $\difsp{\Nbar}{T}[\mu,T=\Tbar]$ one can get the entropy by integration using eq.~\ref{eq:maxwell}.
	
	In the following section we extract the entropy of the one- and two-carrier charge states of the QD using both methods.

	\section{\label{sec:onetwocarrier} One- \& Two-carrier Entropy}

	Figures~\ref{fig:entropy} (b)-(d) show the measured charge detector current and extracted entropy for the $0\to 1$--transition.
	The resulting traces are obtained by scanning the $0 \to 1$--transition several times and averaging the individual traces.
	As the macrostate $\Nbar = 0$ is realized by only one microstate, the vacuum state $\ket{0}$, it is non-degenerate. We define its entropy to be $S_0=0$. Gradually increasing the occupation number to $\Nbar = 1$ by increasing $\mu$, which is effectively done by lowering $\mu_1$ with the plunger gate PG, the entropy reaches a maximum of $\kB\ln(3)$ before settling at $\kB \ln(2)$. The entropy change for the $0\to 1$--transition is therefore $\dl.Delta.{S}_{0\to 1} = S_1 - S_0 = \kB \ln(2)$. This indicates that the ground state of the macrostate $\Nbar=1$ is realized by two microstates. This is in agreement with observations recently made by performing counting statistics measurements on BLG QDs~\cite{duprezSpinvalleyLockedExcited2024}: The ground state of the $N=1$ charge state is the Kramers pair $\Kdown$, $\Kup$, hence two-fold degenerate at zero magnetic field. The entropy peak at $\kB \ln(3)$, where the chemical potential $\mu$ equals the addition energy $\mu_1$ to add the first carrier in its ground state, corresponds to the QD being equally likely in any of the following three microstates: $\ket{0}$, $\Kdown$, $\Kup$. Method A and B give results that agree with each other (see Fig.~\ref{fig:heating}(b)~\cite{supplemental2025} for a comparison of these methods also over different electronic temperature ranges).

	For the $1\to 2$--transition shown in Figs.~\ref{fig:entropy} (e)-(g), the entropy for the macrostate $\Nbar = 1$ is given by $S_1 = \kB \ln(2)$. Gradually increasing the occupation number to $\Nbar = 2$ by further increasing $\mu$, leads to an entropy maximum of $\kB \ln(3)$ before settling at $S_2 = 0$. The entropy change for the $1\to 2$--transition is therefore $\dl.Delta.{S}_{1\to 2} = -\kB\ln(2)$. This lets us conclude that the ground state of the macrostate $\Nbar=2$ is realized by only one microstate. The result that the ground state of the $N=2$ charge state is nondegenerate is a surprise as previously studied devices hinted towards a threefold degeneracy: a valley-singlet spin-triplet ground state with a three-fold spin-degeneracy, a conclusion reached by extrapolating finite bias measurements performed at higher magnetic fields to zero field~\cite{kurzmannExcitedStatesBilayer2019, mollerProbingTwoElectronMultiplets2021}.
	The entropy peak at $\kB \ln(3)$, where the chemical potential $\mu$ equals the addition energy $\mu_2$ to add the second carrier in its ground state, corresponds to the QD being equally likely in any of the following three microstates: $\Kdown$, $\Kup$ and $\GStwo$, where the latter ket denotes the unknown ground state of the $N=2$ charge state.
	The extracted entropies are independent from the electronic temperatures (see Fig.~\ref{fig:heating}(b)) and the exact barrier gate voltage tunings (see Fig.~\ref{fig:gates} (a) and (b)~\cite{supplemental2025}).
	In the next section, we compare entropy measurements and finite bias spectroscopy in an out-of-plane magnetic field with each other in order to gain more insights into the nature of this unknown two particle ground state.



	\section{\label{sec:magnet} Entropy in Magnetic Field}

	\begin{figure*}[t!]
		\includegraphics{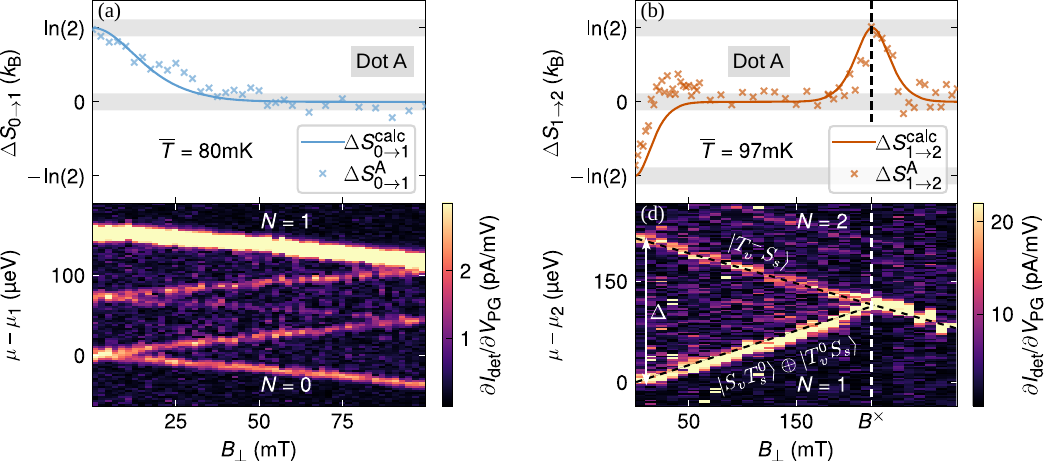}
		\caption{\label{fig:magnet}
			Evolution of the measured entropy change in out-of-plane magnetic field together with the calculated entropy change (solid line) for the $0\to 1$ (a) and $1\to 2$ transition (b). for the $0\to 1$ and $1\to 2$ transitions. Finite bias spectroscopy measurements in out-of-plane magnetic field performed for the $0\to 1$ (c) and $1\to 2$ transition (d). The states are identified by their behavior in out-of-plane magnetic field.
		}
	\end{figure*}
	
	Figures~\ref{fig:magnet}(a), (b) show the entropy changes $\dl.Delta.{S}_{0\to 1}^\mathrm{A}$ and $\dl.Delta.{S}_{1\to 2}^\mathrm{A}$ for the $0\to 1$-- and $1\to 2$--transition, respectively, as a function of out-of-plane magnetic field $\Bper$. For the $0\to 1$--transition in Fig.~\ref{fig:magnet}(a) the entropy change decreases from $\kB \ln(2)$ to $0$ upon increasing $\Bper$. This is in agreement with the Kramers pair $\Kdown$, $\Kup$ being the one carrier ground state at zero magnetic field: The degeneracy is gradually lifted with increasing $\Bper$ due to the spin- and valley-Zeeman effect, whose strengths are characterized by the Land\'e $g$-factors $g_s$ and $g_v$, respectively. When the Zeeman splitting $(g_v+g_s)\muB\Bper$ between states $\Kdown$ and $\Kup$ exceeds the thermal energy $\kB\Tbar$, the degeneracy appears as fully lifted with the ground state being $\Kdown$. This is also observed for the $1\to 2$--transition in Fig.~\ref{fig:magnet}(b) at magnetic fields below $\SI{100}{\milli\tesla}$: upon increasing $\Bper$ the entropy change increases from $-\kB\ln(2)$ to $0$ as the degeneracy of the one carrier state is lifted. Since the low-field region up to $\SI{100}{\milli\tesla}$ in Fig.~\ref{fig:magnet}(b) is fully explained with the behavior of the Kramers pair $\Kdown$, $\Kup$ in $\Bper$, we conclude that the ground state $\GStwo$ of the two-carrier charge state does not depend on $\Bper$. This motivates the chosen notation, indicating that it can be best represented as a superposition ``$\oplus$'' of a valley-singlet spin-triplet state $\ket{S_v T_s^0}$ and a valley-triplet spin-singlet state $\ket{T_v^0 S_s}$, whose energy does not depend on magnetic field.
	Increasing $\Bper$ beyond the low field region, a peak in entropy change of $\kB\ln(2)$ is observed at $\Bcross = \SI{220}{\milli\tesla}$. The ground state degeneracy of the one-carrier charge state is fully lifted at this field, which lets us conclude that this peak is due to a ground state crossing appearing for the two-carrier charge state.
	
	To support these claims we perform finite bias spectroscopy on the QD in an out-of-plane magnetic field. For that we operate the QD in a different regime where it is symmetrically coupled to two leads by adjusting the voltages of the barrier gates LB and RB (see Fig.~\ref{fig:device}~\cite{supplemental2025}). This is different from the regime in which the entropy measurements are performed, where the QD is only coupled to one lead, with the coupling controlled by RB. A source-drain bias $\VSD$ is applied as well as a voltage modulation $\partial\Vpg$ to the plunger gate PG in order to measure the transconductance $\difsp{\Idet}{\Vpg}$. By carefully tuning the coupling to the leads, an excited state in the bias window increases the average occupation number of the dot, which expresses itself in the appearance of a peak in the transconductance. 
	
	Figures~\ref{fig:magnet}(c), (d) show the transconductance measurements for the $0\to 1$-- and $1\to 2$--transition in magnetic field, respectively. The peaks corresponding to the relevant states are labelled accordingly. For the $0\to 1$--transition in Fig.~\ref{fig:magnet} (c) the splitting of the Kramers pair $\Kdown$, $\Kup$ in out-of-plane magnetic field can be seen. The extended data set in Fig.~\ref{fig:onehole}~\cite{supplemental2025} allows us to extract the Kane-Mele spin-orbit gap~\cite{kurzmannExcitedStatesBilayer2019, mollerProbingTwoElectronMultiplets2021} $\DSO = \SI{75}{\micro\electronvolt}$ as well as the valley and spin $g$-factors $g_v^{(1)} = 13.6$ and $g_s = 2.0$ (see Tab.~\ref{tab:oneparams}) for the first carrier. The $g$-factors together with the temperature $\Tbar$ allow us to calculate the evolution of the entropy change $\dl.Delta.{S}_{0\to 1}^\mathrm{calc}$ (eq.~\ref{eq:dS01}~\cite{supplemental2025}) which is plotted in Fig.~\ref{fig:magnet} (a), showing agreement with the measured entropy change.
	
	For the $1\to 2$--transition in Fig.~\ref{fig:magnet}(d) we can see that the transconductance peak corresponding to $\GStwo$ appears to shift in $\Bper$. This is due to the fact that the addition energy $\mu_2$ is changed, and this quantity also depends on the energy of the ground state of the one-carrier charge state. Hence, from the slope of $\GStwo$ one can extract the valley $g$-factor for the first carrier $g_v^{(1)} = 15.5$ (see Tab.~\ref{tab:twoparams}) assuming $g_s = 2$, in contrast to $13.6$ measured for the $0\to 1$--transition (see Tab.~\ref{tab:oneparams}). The difference is due to the confinement dependence of the valley $g$-factor~\cite{tongTunableValleySplitting2021} which changes as the barrier gate voltages on LB and RB are retuned between measurements of the $0\to 1$-- and $1\to 2$--transition.
	There appears an excited state with an energy gap of $\Delta = \SI{205}{\micro\electronvolt}$ at zero magnetic field, whose energy is lowered rapidely as magnetic field increases, causing the ground state crossing at $\Bcross$. The latter fact leads us to conclude that this is a valley-triplet spin-singlet state $\ket{T_v^- S_s}$ with a valley $g$-factor of $g_v^{(2)} = 16$. An extended data set is shown in Fig.~\ref{fig:twohole}~\cite{supplemental2025}. With these $g$-factors and the temperature $\Tbar$ we can calculate the theoretically expected entropy change $\dl.Delta.{S}_{1\to 2}^\mathrm{calc}$ (eq.~\ref{eq:dS12}~\cite{supplemental2025}) in Fig.~\ref{fig:magnet}(b) showing good agreement with the measurement.

	\section{\label{sec:discussion} Discussion \& Outlook}

	\begin{figure}[t!]
		\includegraphics{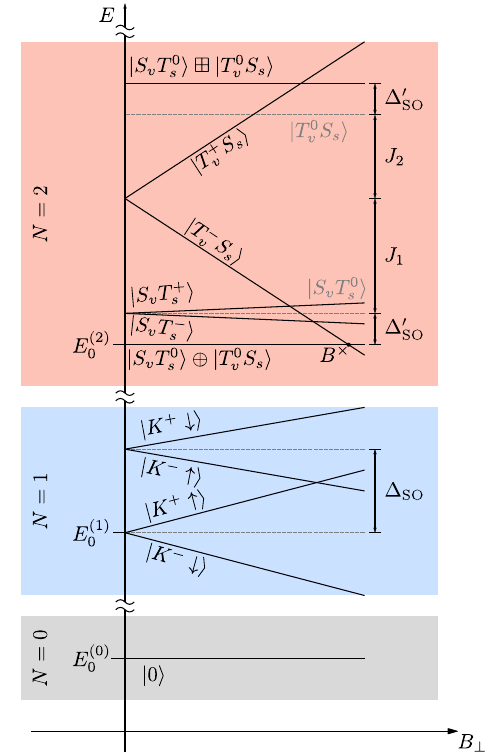} 
		\caption{\label{fig:spectrum}
			Excited state spectra for the one-carrier and two-carrier charge state in an out-of-plane magnetic field including the effect of Kane-Mele type spin--orbit interaction for the $N=2$ charge state. The ground state $\ket{S_v T_s^0} \oplus \ket{T_v^0 S_s}$ is separated from the spin-triplet states $\ket{S_v T_s^+}$, $\ket{S_v T_s^-}$ by an energy gap $\DSO'$.
			The ground state crossing is due to the valley-triplet state $\ket{T_v^- S_s}$ and appears at an out-of-plane magnetic field $\Bcross$.
		}
	\end{figure}

	Measuring entropy allows us to directly determine the ground state degeneracy of charge states in a BLG QD. This approach gives insights beyond conventional transport spectroscopy techniques where degeneracies are detected in an indirect way by the energy levels splitting in magnetic field. The extracted entropy change for the $0\to 1$--transition is consistent with the excited state spectrum measured with finite bias spectroscopy and also previous findings~\cite{kurzmannExcitedStatesBilayer2019, mollerProbingTwoElectronMultiplets2021}: The ground state of the one-carrier charge state has a two-fold degeneracy (Kramers pair $\ket{K^-\downarrow}$, $\ket{K^+\uparrow}$) lifted in out-of-plane magnetic field due to the valley- and spin-Zeeman effect. For the $1\to 2$--transition the extracted entropy change reveals a nondegenerate ground state for the two-carrier charge state at zero magnetic field. This has not been observed before. Previous studies hinted towards a three-fold degenerate ground state given by a valley-singlet spin-triplet state~\cite{kurzmannExcitedStatesBilayer2019, mollerProbingTwoElectronMultiplets2021}.
	We attribute this finding to the effect of the Kane-Mele type spin--orbit interaction~\cite{kaneQuantumSpinHall2005, konschuhTheorySpinorbitCoupling2012}. For the two-carrier charge state, the Kane-Mele type spin--orbit interaction leads to a coupling of the valley-singlet spin-triplet state $\ket{S_v T_s^0}$ with the valley-triplet spin-singlet state $\ket{T_v^0 S_s}$ (see model Hamiltonian in eq.~\ref{eq:hamtwo}~\cite{supplemental2025}). This results in the level spectrum shown in Fig.~\ref{fig:spectrum}. The three-fold degeneracy between $\ket{S_v T_s^+}$, $\ket{S_v T_s^-}$ and $\ket{S_v T_s^0}$ is lifted and a new ground state $\ket{S_v T_s^0} \oplus \ket{T_v^0 S_s}$ emerges, which is a superposition of $\ket{S_v T_s^0}$ and $\ket{T_v^0 S_s}$ (the coefficients are determined by eq.~\ref{eq:theta}~\cite{supplemental2025}). Its energy is lowered by
	\begin{equation}
		\label{eq:DSOtwo}
		\DSO' = \sqrt{\left(\frac{J_1 + J_2}{2}\right)^2 + \DSO^2} - \frac{J_1+J_2}{2}
	\end{equation}
	compared to the states $\ket{S_v T_s^-}$, $\ket{S_v T_s^+}$. Here, $\DSO$ denotes the Kane-Mele spin-orbit gap, and $J_1$, $J_2$ are energy splittings induced by exchange interaction~\cite{kurzmannExcitedStatesBilayer2019}. The extended data set on finite bias spectroscopy in magnetic field (Fig.~\ref{fig:twohole}(b), (c)~\cite{supplemental2025}) indeed shows states as shown in Fig.~\ref{fig:spectrum}. The energy splittings extracted from it are given by $\DSO' = \SI{45}{\micro\electronvolt}$ and $J_1 = \SI{160}{\micro\electronvolt}$. Assuming $J_1 = J_2$ this results in $\DSO'^{\mathrm{calc}} = \SI{20}{\micro\electronvolt}$, a value within a factor of two compared to the measured splitting. Eq.~\ref{eq:DSOtwo} implies that the exceptionally small $J_1$ enables us to resolve $\DSO'$.
    Previously reported excited state spectroscopy data lacks resolution in energy~\cite{kurzmannKondoEffectSpin2021,mollerProbingTwoElectronMultiplets2021} or/and reported exchange splittings renders the expected gap undetectable~\cite{kurzmannExcitedStatesBilayer2019}: $J_1 = \SI{350}{\micro\electronvolt}$ implies $\DSO'^{\mathrm{calc}} \approx \SI{10}{\micro\electronvolt}$.

	According to Ref.~\cite{knotheQuartetStatesTwoelectron2020}, the intervalley-exchange interaction is determined by the product of a BLG specific constant $g_\perp$ and an integral $W_{\alpha,\beta}$ over products of two-particle orbital wave functions. This implies that the exchange interaction depends on the spatial extent of the orbital wave function. The theoretical estimates of $J_1, J_2$ of the order of $\SI{100}{\micro\electronvolt}$ in~\cite{knotheQuartetStatesTwoelectron2020} agree with all experimental findings including those in our paper. Based on this theory, $J_1, J_2$ depend on the spatial shape of the wave functions (and thereby in principle on device geometry and disorder potential), and weakly on the displacement field.
	This new two-carrier ground state is also observed in Dot B, which is a pn-junction defined electron type QD at a location $250-\SI{300}{\nano\meter}$ away from Dot A operated at a different displacement field (see Figs.~\ref{fig:electronstates},~\ref{fig:1stelectron} and~\ref{fig:2ndelectron}~\cite{supplemental2025}).
	This raises further interesting questions regarding the tunability of $\DSO'$ and what consequences this bears for building singlet triplet qubits~\cite{pettaCoherentManipulationCoupled2005} in BLG QDs.
    
	In conclusion, we demonstrate the feasibility of Maxwell relation based entropy measurements in gate-defined BLG QDs. For the one-carrier regime it gives the same result as excited state spectroscopy. In the two-carrier regime we uncover a previously unobserved ground state which we attribute to the effect of Kane-Mele spin-orbit interaction and provide a single-particle model which is capable of producing the observed energy spectrum. We also demonstrate that this new ground state appears at different temperatures, displacement fields, and QDs of different carrier type and location.
	Beyond quantum dots, our work generalizes and extends the Maxwell relation-based entropy measurements to samples fabricated in novel 2D vdW heterostructures and shows that sophisticated device designs can be implemented in these material systems, paving the way to probe various exotic states in BLG devices. Examples of these are the (multichannel-) Kondo effect~\cite{kurzmannKondoEffectSpin2021, potokObservationTwochannelKondo2007} or quantum spin chains with fractional or integer spins~\cite{haldaneNonlinearFieldTheory1983,haldaneContinuumDynamics1D1983}.

	\begin{acknowledgments}
	We appreciate the discussions with J. Folk, E. Sela, and Y. Meir concerning entropy in mesoscopic systems. We thank P. Märki, T. Bähler, and FIRST staff for technical support. We acknowledge support from the European Graphene Flagship Core3 Project, Swiss National Science Foundation via NCCR Quantum Science, and H2020 European Research Council (ERC) Synergy Grant under Grant Agreement 95154. K.W. and T.T. acknowledge support from the JSPS KAKENHI (Grant Numbers 20H00354 and 23H02052) and World Premier International Research Center Initiative (WPI), MEXT, Japan.
\end{acknowledgments}

	\bibliography{bibliography} 
	\clearpage

	\appendix

	\renewcommand{\theequation}{S\arabic{equation}} 
	\setcounter{equation}{0} 

	\renewcommand{\thefigure}{S\arabic{figure}} 
	\setcounter{figure}{0} 

	\begin{onecolumngrid}
	\section{Sample Design \& Operation}
		\label{app:sample}
		\begin{figure*}[h!]
			\includegraphics{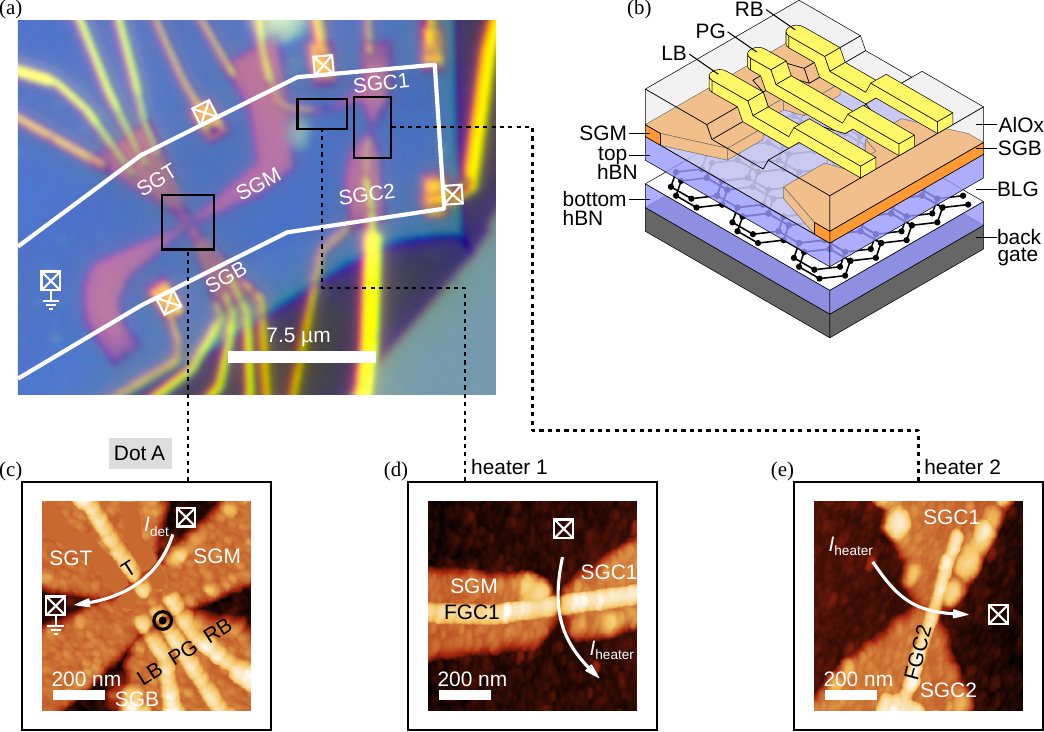}
			\caption{\label{fig:device}(a) Optical microscope image of the used sample. The BLG flake is outlined in white. The applied back gate voltage is negative, rendering the BLG p-type/hole conducting. The split gate pairs SGM, SGB and SGM, SGT are used to electrostatically define a conducting channel. (b) Schematic of the VdW stack structure in the region where the QD is defined: BLG is encapsulated by hexagonal boron nitride (hBN) surrounded by a global graphite back gate ($\mathrm{BG}$) and with two layers of gold top gates: split gates SGM, SGB to form a one-dimensional channel, and finger gates to form the QD (plunger gate PG and tunneling barrier gates LB, RB). The top gate layers are separated by aluminum oxide ($\mathrm{AlOx}$) as a dielectric. The displacement field is roughly $\SI{-0.46}{\volt\per\nano\meter}$ resulting in a band gap of approximately $\SI{50}{\milli\electronvolt}$\cite{ickingTransportSpectroscopyUltraclean2022}. The device is operated in the hole conduction regime. The barriers are tuned such that there is no coupling to the left lead. AFM images of the (c) QD (black circles) formed with finger gates LB, PG, RB and the charge detector tuned with finger gate T, and the (d) (e) heater gate structures to form the ohmic constrictions, which are formed with split gates SGM, SGC1, SGC2 and finger gates FGC1 and FGC2.
			}
		\end{figure*}

		\begin{figure}[h!]
			\includegraphics{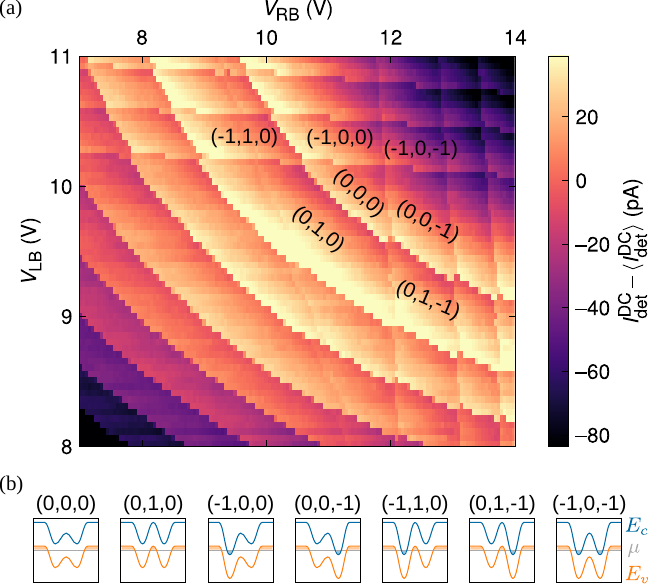}
			\caption{\label{fig:operation}
				(a) Charge detector current as a function of left and right barrier voltage. The plunger gate voltage is fixed to $\SI{3.25}{\volt}$. Three types of resonances, corresponding to three types of QDs are observed:
				1. pn-junction defined electron QD under LB corresponding to the horizontal transconductance peaks.
				2. barrier defined hole QD between LB and RB corresponding to the bent transconductance peaks. This is the QD used for the entropy measurements.
				3. pn-junction defined electron QD under RB corresponding to the vertical transconductance peaks. The number triplet labels the occupation number of the respective QDS (1., 2. 3.). (b) Diagrams showing the band bending induced by the finger gates LB, PG, RB in order to form the different types of QDs for a few exemplary cases. QDs form where the conduction band edge $E_c$ or valence band edge $E_v$ crosses the chemical potential $\mu$ in energy.
			}
		\end{figure}
		
		\begin{figure}[h!]
			\includegraphics{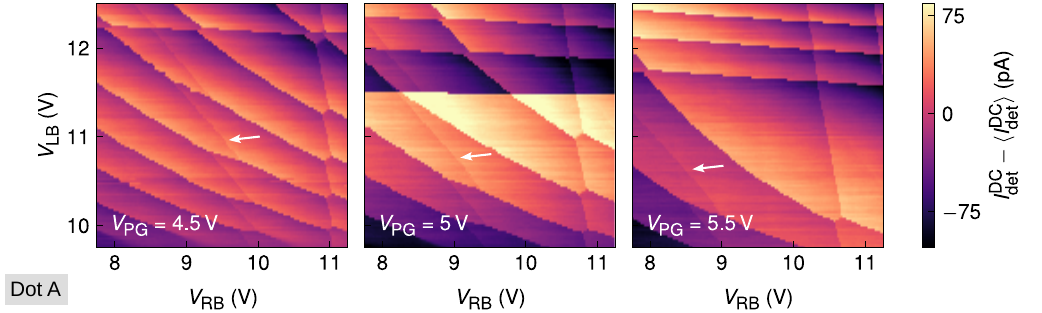}
			\caption{\label{fig:defectmaps}
				(a) Charge detector current as a function of left and right barrier voltage for different plunger gate voltages. The types of resonances are the same as in Fig.~\ref{fig:operation}. The white arrow indicates a defect/spurious QD. By tuning the gate voltages $\VL$, $\VR$, $V_\mathrm{PG}$ we make sure that the investigated QD is off-resonance with any defects/spurious QDs and measurements are not affected.
			}
		\end{figure}

\end{onecolumngrid} \clearpage
	\begin{onecolumngrid}

	\section{\label{app:heating} Heating Calibration}

		\begin{figure*}[h!]
			\includegraphics{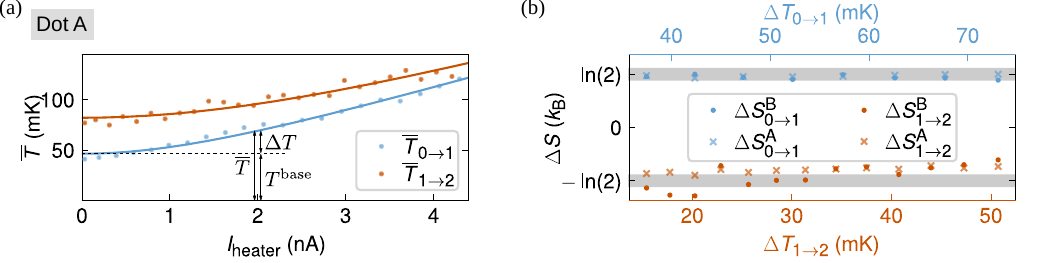}
			\caption{\label{fig:heating}
            (a) shows the mean temperature of the hole bath forming the reservoir as a function of the heater current. Temperatures for individual heater currents are obtained by tuning the QD into the thermally broadened regime and fitting eq.~\ref{eq:detDC} to $\Idet^\mathrm{DC}$. The evolution of $\Tbar$ in $\Iheater$ follows eq.~\ref{eq:Thyp} as can be seen from the fit. The experiments for the $0\to 1$-- and $1\to 2$-- transition were carried out at different hole reservoir base temperatures were $\Iheater = 0$, namely $T^\mathrm{base}_{0\to 1} = \SI{45}{\milli\kelvin}$ and $T^\mathrm{base}_{1\to 2} = \SI{82}{\milli\kelvin}$. The temperature modulations are extracted from the calibration by via $\dl.Delta.{T}_{0\to 1}=\Tbar_{0\to 1} - T^\mathrm{base}_{0\to 1}$ and $\dl.Delta.{T}_{1\to 2}=\Tbar_{1\to 2} - T^\mathrm{base}_{1\to 2}$. (b) shows the entropy changes obtained by  applying methods A and B plotted against the temperature modulation. The extracted entropy changes remain constant over the given temperature modulation range. Both methods show reasonable agreement with each other. 
			}
		\end{figure*}

        \begin{figure*}[h!]
			\includegraphics{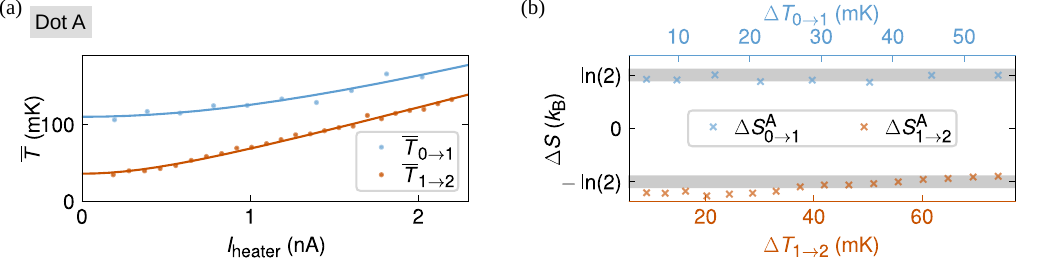}
			\caption{\label{fig:differentbase}
				(a) and (b) show the same measurements as Fig.~\ref{fig:heating} carried out at different base temperatures of the hole bath, namely $T^\mathrm{base}_{0\to 1} = \SI{110}{\milli\kelvin}$ and $T^\mathrm{base}_{1\to 2} = \SI{35}{\milli\kelvin}$. The obtained entropy changes in (b) remain independent of the temperature modulations $\dl.Delta.{T}_{0\to 1} = \Tbar_{0\to 1} - T^\mathrm{base}_{0\to 1}$ and $\dl.Delta.{T}_{1\to 2} = \Tbar_{1\to 2} - T^\mathrm{base}_{1\to 2}$.
			}
		\end{figure*}
\end{onecolumngrid} \clearpage
	\begin{onecolumngrid}

	\section{\label{app:averaging} Averaging Procedure}
		
		\begin{figure*}[h!]
			\includegraphics{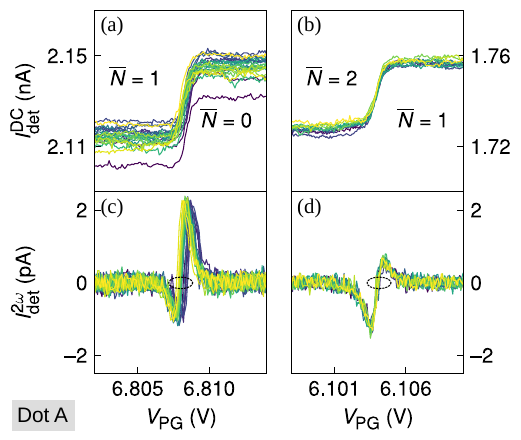}
			\caption{\label{fig:rawdata} Raw data traces of the DC components (a), (b) and second harmonic components (c), (d) of the charge detector current for the $0\to 1$-- and $1\to 2$-- transition, respectively. The data traces for the respective transitions are aligned with respect to the zero-crossing of the second harmonic component (dashed ellipses in (c) and (d)). The aligned traces are then averaged resulting in the traces shown in Fig.~\ref{fig:entropy}.
			}
		\end{figure*}
	
\end{onecolumngrid} \clearpage
	\begin{onecolumngrid}

	\section{\label{app:gates} Robustness with respect to barrier voltage configuration and CD bias}

		\begin{figure*}[h!]
			\includegraphics{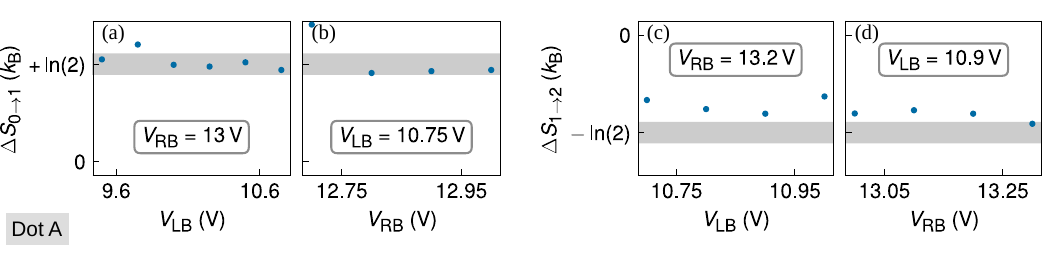}
			\caption{\label{fig:gates} Extracted entropy changes for the $0\to 1$--transition (a) and the $1\to 2$--transition (b) for different barrier gate configurations. The extracted entropy changes show no dependence on the barrier voltage configuration over the range of a few $\SI{100}{\milli\volt}$.}

		\end{figure*}

        \begin{figure*}[h!]
		\includegraphics{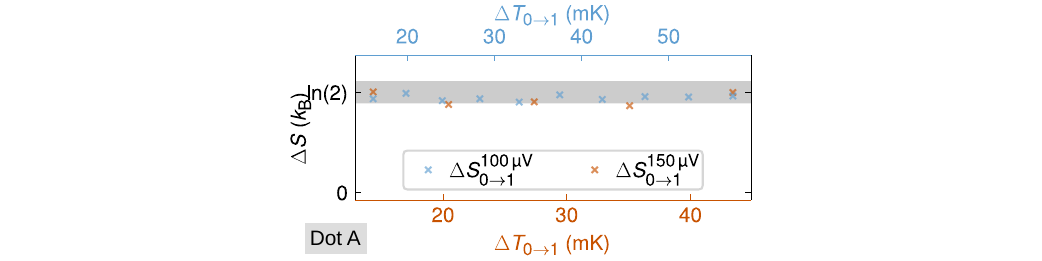}
		\caption{
			Measurements of the entropy change as a function of the temperature modulation for the $0\to 1$--transition, carrier out at the two charge detector bias voltages $\SI{100}{\micro\volt}$ and $\SI{150}{\micro\volt}$. In this regime, the extracted entropy values are independent of the charge detector bias.
			}
	\end{figure*}

\end{onecolumngrid} \clearpage
	\def\Ham{\ensuremath{H}}
\def\Hil{\ensuremath{\mathcal{H}}}

\def\Orb{\ensuremath{\Delta_\mathrm{Orb}}}

\def\Kp{\ensuremath{{^+}}}
\def\Km{\ensuremath{{^-}}}

\def\Sd{\ensuremath{\downarrow}}
\def\Su{\ensuremath{\uparrow}}

\def\Vc{\ensuremath{V_\mathrm{C}}}

\begin{onecolumngrid}
	\section{Excited state spectra for one \& two carriers}
	\label{app:spectrum}
	Here, we put forward a model that captures the effect of Kane-Mele type spin-orbit coupling (SOC), as present in BLG, on the energy spectrum of the second charge carrier state. The model does not capture any interaction-effects potentially changing the Kane-Mele spin-orbit gap $\DSO$. SOC in bilayer graphene is intrinsically present or externally induced. The intrinsic type of SOC, commonly termed Kane-Mele (KM) SOC~\cite{kaneQuantumSpinHall2005} is of the form
	\begin{equation}
		\Ham_\mathrm{KM} = \frac{\Delta_\mathrm{KM}}{2} \sigma_z \tau_z s_z
	\end{equation}
	where the Pauli matrices $\sigma_z$, $\tau_z$ and $s_z$ operate in the sublattice-, valley- and spin-subspace, respectively. KM SOC in bilayer graphene arises essentially from the same mechanism as in monolayer graphene~\cite{konschuhTheorySpinorbitCoupling2012}, which in turn is attributed to the nominally unoccupied $d$-orbitals in graphene~\cite{konschuhTightbindingTheorySpinorbit2010}. $\Delta_\mathrm{KM}$ is independent from the applied displacement field~\cite{konschuhTheorySpinorbitCoupling2012}. \newline
	The extrinsic type of SOC, commonly termed Bychkov-Rashba (BR) SOC~\cite{bychkovOscillatoryEffectsMagnetic1984a} is of the form~\cite{kaneQuantumSpinHall2005}
	\begin{equation}
		\Ham_\mathrm{BR} = \frac{\Delta_\mathrm{BR}}{2} \left(\sigma_x \tau_z s_y - \sigma_y s_x\right)
	\end{equation}
	It stems from the broken inversion symmetry between the graphene layers, e.g. due to an applied displacement field. $\Delta_\mathrm{BR}$ is expected to depend linearly on the applied displacement field~\cite{konschuhTheorySpinorbitCoupling2012}. However, the SO gap observed in the investigated QDs proofs to be independet of the displacement field, carrier type and confinement (tunneling barrier defined VS pn-junction defined), and at a value of roughly $\SI{80}{\micro\electronvolt}$ it is also in line with previous observations~\cite{kurzmannKondoEffectSpin2021,mollerProbingTwoElectronMultiplets2021, duprezSpinvalleyLockedExcited2024}. Hence, we conclude that the observed spin-orbit gap is of the KM type.

	The two-particle quantum states of the QD exist in the Hilbert space
	\begin{equation}
		\Hil = \Hil_{\bm{r}}^{(1)} \otimes \Hil_{\bm{r}}^{(2)} \otimes \Hil_v^{(1)} \otimes \Hil_v^{(2)} \otimes \Hil_s^{(1)} \otimes \Hil_s^{(2)}
	\end{equation}
	where indices $\bm{r}$, $v$ and $s$ denote the orbital-, valley- and spin-subspace of particle (1) and (2), respectively. The single-particle orbital spacing $\Orb$ is of the order of $\SI{1}{\milli\electronvolt}$, much larger than $\DSO \approx \SI{80}{\micro\electronvolt}$. Hence, the first few carriers are assumed to always occupy the ground state orbital, allowing us to focus only on the valley- and spin-subspace. In the single-particle basis $\left\{\ket{K^-} \otimes \ket{\downarrow}, \ket{K^+}\otimes\ket{\uparrow}, \ket{K^-} \otimes \ket{\uparrow}, \ket{K^+} \otimes \ket{\downarrow} \right\}$, which we abbreviate as $\left\{ \ket{\Km\Sd}, \ket{\Kp\Su}, \ket{\Km\Su}, \ket{\Kp\Sd} \right\}$, the spin-orbit hamiltonian for one particle is given by
	\begin{align}
		\Ham_\mathrm{SO}^{(1)}
		&= \frac{\DSO}{2}
		\begin{pNiceMatrix}[first-row, nullify-dots, columns-width=auto, columns-type=r]
			\ket{K^-\Sd} & \ket{K^+\Su} & \ket{K^-\Su} & \ket{K^+\Sd} \\
			-1 &  0 & 0 & 0 \\
			0 & -1 & 0 & 0 \\
			0 &  0 & 1 & 0 \\
			0 &  0 & 0 & 1
		\end{pNiceMatrix}
	\end{align}
	and the Zeeman hamiltonian is
	\begin{align}
		H_\mathrm{Z}^{(1)}
		&= \frac{\muB\Bper}{2}
		\begin{pNiceMatrix}[first-row, nullify-dots, columns-width=auto]
			\ket{K^-\Sd} & \ket{K^+\Su} & \ket{K^-\Su} & \ket{K^+\Sd} \\
			-(g^{(1)}_v+g_s) &    0    &      0     &    0   \\
			0     & g^{(1)}_v+g_s &      0     &    0   \\
			0     &    0    & -(g^{(1)}_v-g_s) &    0   \\
			0     &    0    &      0     & g^{(1)}_v-g_s
		\end{pNiceMatrix}
	\end{align}
	The hamiltonian $H_\mathrm{ES}^{(1)} = H_\mathrm{SO}^{(1)} + H_\mathrm{Z}^{(1)} $ contains the  low-energy excitation spectrum for the single-particle regime and is given by
	\begin{equation}
		\label{eq:hamone}
		H_\mathrm{ES}^{(1)} = \frac{1}{2}
		\begin{pNiceMatrix}[first-row, nullify-dots, columns-width=auto]
			\ket{K^-\Sd} & \ket{K^+\Su} & \ket{K^-\Su} & \ket{K^+\Sd} \\
			-\DSO - (g^{(1)}_v+g_s)\muB\Bper &    0    &      0     &    0   \\
			0     & -\DSO + (g^{(1)}_v+g_s)\muB\Bper &      0     &    0   \\
			0     &    0    & \DSO - (g^{(1)}_v-g_s)\muB\Bper &    0   \\
			0     &    0    &      0     & \DSO - (g^{(1)}_v-g_s)\muB\Bper
		\end{pNiceMatrix}
	\end{equation}

	\clearpage
	\setcounter{MaxMatrixCols}{20}
	\begin{sideways}
		\parbox{\textheight}{
			The SO hamiltonian for two particles $H^{(1)\otimes(2)}_\mathrm{SO} = H_\mathrm{SO}^{(1)} \oplus H_\mathrm{SO}^{(2)}$ is then given by
			\begin{equation}
				H_\mathrm{SO}^{(1)\otimes(2)} = \DSO \begin{pNiceMatrix}[first-row, nullify-dots,columns-type=c]
					\ket{\Km\Km\Sd\Sd} & \ket{\Km\Kp\Sd\Su} & \ket{\Km\Km\Sd\Su} & \ket{\Km\Kp\Sd\Sd} &
					\ket{\Kp\Km\Su\Sd} & \ket{\Kp\Kp\Su\Su} & \ket{\Kp\Km\Su\Su} & \ket{\Kp\Kp\Su\Sd} &
					\ket{\Km\Km\Su\Sd} & \ket{\Km\Kp\Su\Su} & \ket{\Km\Km\Su\Su} & \ket{\Km\Kp\Su\Sd} &
					\ket{\Kp\Km\Sd\Sd} & \ket{\Kp\Kp\Sd\Su} & \ket{\Kp\Km\Sd\Su} & \ket{\Kp\Kp\Sd\Sd} \\
					-1 & 0 & 0 & 0 & 0 & 0 & 0 & 0 & 0 & 0 & 0 & 0 & 0 & 0 & 0 & 0 \\
					0 &-1 & 0 & 0 & 0 & 0 & 0 & 0 & 0 & 0 & 0 & 0 & 0 & 0 & 0 & 0 \\
					0 & 0 & 0 & 0 & 0 & 0 & 0 & 0 & 0 & 0 & 0 & 0 & 0 & 0 & 0 & 0 \\
					0 & 0 & 0 & 0 & 0 & 0 & 0 & 0 & 0 & 0 & 0 & 0 & 0 & 0 & 0 & 0 \\
					0 & 0 & 0 & 0 &-1 & 0 & 0 & 0 & 0 & 0 & 0 & 0 & 0 & 0 & 0 & 0 \\
					0 & 0 & 0 & 0 & 0 &-1 & 0 & 0 & 0 & 0 & 0 & 0 & 0 & 0 & 0 & 0 \\
					0 & 0 & 0 & 0 & 0 & 0 & 0 & 0 & 0 & 0 & 0 & 0 & 0 & 0 & 0 & 0 \\
					0 & 0 & 0 & 0 & 0 & 0 & 0 & 0 & 0 & 0 & 0 & 0 & 0 & 0 & 0 & 0 \\
					0 & 0 & 0 & 0 & 0 & 0 & 0 & 0 & 0 & 0 & 0 & 0 & 0 & 0 & 0 & 0 \\
					0 & 0 & 0 & 0 & 0 & 0 & 0 & 0 & 0 & 0 & 0 & 0 & 0 & 0 & 0 & 0 \\
					0 & 0 & 0 & 0 & 0 & 0 & 0 & 0 & 0 & 0 &+1 & 0 & 0 & 0 & 0 & 0 \\
					0 & 0 & 0 & 0 & 0 & 0 & 0 & 0 & 0 & 0 & 0 &+1 & 0 & 0 & 0 & 0 \\
					0 & 0 & 0 & 0 & 0 & 0 & 0 & 0 & 0 & 0 & 0 & 0 & 0 & 0 & 0 & 0 \\
					0 & 0 & 0 & 0 & 0 & 0 & 0 & 0 & 0 & 0 & 0 & 0 & 0 & 0 & 0 & 0 \\
					0 & 0 & 0 & 0 & 0 & 0 & 0 & 0 & 0 & 0 & 0 & 0 & 0 & 0 &+1 & 0 \\
					0 & 0 & 0 & 0 & 0 & 0 & 0 & 0 & 0 & 0 & 0 & 0 & 0 & 0 & 0 &+1
				\end{pNiceMatrix}
			\end{equation}
			Instead of representing the two-particle Hilbert space as a tensor produkt $\Hil = \Hil^{(1)} \otimes \Hil^{(2)}$, one can also consider it to be direct sum of the subspaces spanned by antisymmetrized and symmetrized states $\Hil = \Hil^{(\mathrm{AS})} \oplus \Hil^{(\mathrm{S})}$. The preferred basis states are given by valley- and spin- singlets and triplets. The unitary matrix $U$ transforming between both sets of basis states is given by
			\begin{equation}
				U = \begin{pNiceMatrix}[first-col, first-row, nullify-dots]
					&\ket{\Km\Km\Sd\Sd} & \ket{\Km\Kp\Sd\Su} & \ket{\Km\Km\Sd\Su} & \ket{\Km\Kp\Sd\Sd} &
					\ket{\Kp\Km\Su\Sd} & \ket{\Kp\Kp\Su\Su} & \ket{\Kp\Km\Su\Su} & \ket{\Kp\Kp\Su\Sd} &
					\ket{\Km\Km\Su\Sd} & \ket{\Km\Kp\Su\Su} & \ket{\Km\Km\Su\Su} & \ket{\Km\Kp\Su\Sd} &
					\ket{\Kp\Km\Sd\Sd} & \ket{\Kp\Kp\Sd\Su} & \ket{\Kp\Km\Sd\Su} & \ket{\Kp\Kp\Sd\Sd} \\
					\bra{S_v T_s^-}   & 0 & 0 & 0 & -\frac{1}{\sqrt{2}} & 0 & 0 & 0 & 0 & 0 & 0 & 0 & 0 & \frac{1}{\sqrt{2}} & 0 & 0 & 0 \\
					\bra{S_v T_s^+}   & 0 & 0 & 0 & 0 & 0 & 0 & \frac{1}{\sqrt{2}} & 0 & 0 & -\frac{1}{\sqrt{2}} & 0 & 0 & 0 & 0 & 0 & 0 \\
					\bra{S_v T_s^0}   & 0 & -\frac{1}{2} & 0 & 0 & \frac{1}{2} & 0 & 0 & 0 & 0 & 0 & 0 & -\frac{1}{2} & 0 & 0 & \frac{1}{2} & 0 \\
					\bra{T_v^- S_s}   & 0 & 0 & -\frac{1}{\sqrt{2}} & 0 & 0 & 0 & 0 & 0 & \frac{1}{\sqrt{2}} & 0 & 0 & 0 & 0 & 0 & 0 & 0 \\
					\bra{T_v^+ S_s}   & 0 & 0 & 0 & 0 & 0 & 0 & 0 & \frac{1}{\sqrt{2}} & 0 & 0 & 0 & 0 & 0 & -\frac{1}{\sqrt{2}} & 0 & 0 \\
					\bra{T_v^0 S_s}   & 0 & -\frac{1}{2} & 0 & 0 & \frac{1}{2} & 0 & 0 & 0 & 0 & 0 & 0 & \frac{1}{2} & 0 & 0 & -\frac{1}{2} & 0 \\
					\bra{S_v S_s}     & 0 & \frac{1}{2} & 0 & 0 & \frac{1}{2} & 0 & 0 & 0 & 0 & 0 & 0 & -\frac{1}{2} & 0 & 0 & -\frac{1}{2} & 0 \\
					\bra{T_v^- T_s^-} & 1 & 0 & 0 & 0 & 0 & 0 & 0 & 0 & 0 & 0 & 0 & 0 & 0 & 0 & 0 & 0 \\
					\bra{T_v^- T_s^+} & 0 & 0 & 0 & 0 & 0 & 0 & 0 & 0 & 0 & 0 & 1 & 0 & 0 & 0 & 0 & 0 \\
					\bra{T_v^- T_s^0} & 0 & 0 & \frac{1}{\sqrt{2}} & 0 & 0 & 0 & 0 & 0 & \frac{1}{\sqrt{2}} & 0 & 0 & 0 & 0 & 0 & 0 & 0 \\
					\bra{T_v^+ T_s^-} & 0 & 0 & 0 & 0 & 0 & 0 & 0 & 0 & 0 & 0 & 0 & 0 & 0 & 0 & 0 & 1 \\
					\bra{T_v^+ T_s^+} & 0 & 0 & 0 & 0 & 0 & 1 & 0 & 0 & 0 & 0 & 0 & 0 & 0 & 0 & 0 & 0 \\
					\bra{T_v^+ T_s^0} & 0 & 0 & 0 & 0 & 0 & 0 & 0 & \frac{1}{\sqrt{2}} & 0 & 0 & 0 & 0 & 0 & \frac{1}{\sqrt{2}} & 0 & 0 \\
					\bra{T_v^0 T_s^-} & 0 & 0 & 0 & \frac{1}{\sqrt{2}} & 0 & 0 & 0 & 0 & 0 & 0 & 0 & 0 & \frac{1}{\sqrt{2}} & 0 & 0 & 0 \\
					\bra{T_v^0 T_s^+} & 0 & 0 & 0 & 0 & 0 & 0 & \frac{1}{\sqrt{2}} & 0 & 0 & \frac{1}{\sqrt{2}} & 0 & 0 & 0 & 0 & 0 & 0 \\
					\bra{T_v^0 T_s^0} & 0 & \frac{1}{2} & 0 & 0 & \frac{1}{2} & 0 & 0 & 0 & 0 & 0 & 0 & \frac{1}{2} & 0 & 0 & \frac{1}{2} & 0 \\
				\end{pNiceMatrix}
			\end{equation}
		}
	\end{sideways}
	\clearpage
	\begin{sideways}
		\parbox{\textheight}{
			By applying the transformation $U$ according to $U H_\mathrm{SO}^{(1)\otimes(2)} U^\dagger$ we arrive at
			\begin{equation}
				H_\mathrm{SO} = \DSO\begin{pNiceMatrix}[first-row, nullify-dots]
					\ket{S_v T_s^-} & \ket{S_v T_s^+} & \ket{S_v T_s^0} & \ket{T_v^- S_s} &
					\ket{T_v^+ S_s} & \ket{T_v^0 S_s} & \ket{S_v S_s} & \ket{T_v^- T_s^-} &
					\ket{T_v^- T_s^+} & \ket{T_v^- T_s^0} & \ket{T_v^+ T_s^-} & \ket{T_v^+ T_s^+} &
					\ket{T_v^+ T_s^0} & \ket{T_v^0 T_s^-} & \ket{T_v^0 T_s^+} & \ket{T_v^0 T_s^0} \\
					0 & 0 & 0 & 0 & 0 & 0 & 0 & 0 & 0 & 0 & 0 & 0 & 0 & 0 & 0 & 0 \\
					0 & 0 & 0 & 0 & 0 & 0 & 0 & 0 & 0 & 0 & 0 & 0 & 0 & 0 & 0 & 0 \\
					0 & 0 & 0 & 0 & 0 &-1 & 0 & 0 & 0 & 0 & 0 & 0 & 0 & 0 & 0 & 0 \\
					0 & 0 & 0 & 0 & 0 & 0 & 0 & 0 & 0 & 0 & 0 & 0 & 0 & 0 & 0 & 0 \\
					0 & 0 & 0 & 0 & 0 & 0 & 0 & 0 & 0 & 0 & 0 & 0 & 0 & 0 & 0 & 0 \\
					0 & 0 &-1 & 0 & 0 & 0 & 0 & 0 & 0 & 0 & 0 & 0 & 0 & 0 & 0 & 0 \\
					0 & 0 & 0 & 0 & 0 & 0 & 0 & 0 & 0 & 0 & 0 & 0 & 0 & 0 & 0 &-1 \\
					0 & 0 & 0 & 0 & 0 & 0 & 0 &-1 & 0 & 0 & 0 & 0 & 0 & 0 & 0 & 0 \\
					0 & 0 & 0 & 0 & 0 & 0 & 0 & 0 & 1 & 0 & 0 & 0 & 0 & 0 & 0 & 0 \\
					0 & 0 & 0 & 0 & 0 & 0 & 0 & 0 & 0 & 0 & 0 & 0 & 0 & 0 & 0 & 0 \\
					0 & 0 & 0 & 0 & 0 & 0 & 0 & 0 & 0 & 0 & 1 & 0 & 0 & 0 & 0 & 0 \\
					0 & 0 & 0 & 0 & 0 & 0 & 0 & 0 & 0 & 0 & 0 &-1 & 0 & 0 & 0 & 0 \\
					0 & 0 & 0 & 0 & 0 & 0 & 0 & 0 & 0 & 0 & 0 & 0 & 0 & 0 & 0 & 0 \\
					0 & 0 & 0 & 0 & 0 & 0 & 0 & 0 & 0 & 0 & 0 & 0 & 0 & 0 & 0 & 0 \\
					0 & 0 & 0 & 0 & 0 & 0 & 0 & 0 & 0 & 0 & 0 & 0 & 0 & 0 & 0 & 0 \\
					0 & 0 & 0 & 0 & 0 & 0 &-1 & 0 & 0 & 0 & 0 & 0 & 0 & 0 & 0 & 0
				\end{pNiceMatrix}
			\end{equation}
			The eigenvalues are $-\DSO: \frac{1}{\sqrt{2}} (\ket{S_vT_s^0} + \ket{T_v^0 S_s}), \frac{1}{\sqrt{2}} (\ket{S_vS_s} + \ket{T_v^0 T_v^0}), \ket{T_v^-T_s^-}, \ket{T_v^+T_s^+}$,
			$\DSO: \frac{1}{\sqrt{2}} (-\ket{S_vT_s^0} + \ket{T_v^0 S_s}), \frac{1}{\sqrt{2}} (\ket{S_vS_s} - \ket{T_v^0 T_v^0}), \ket{T_v^-T_s^+}, \ket{T_v^+T_s^-}$, $0:\ket{S_vT_s^\pm}, \ket{T_v^\pm S_s}, \ket{T_v^\pm T_s^0}, \ket{T_v^0 T_s^\pm}$. As the valley states $S_v$, $T_v^\mp$, $T_v^0$ The Zeeman hamiltonian is given by
			\begin{equation}
				H_\mathrm{Z} = \muB\Bper\begin{pNiceMatrix}[first-row, nullify-dots]
					\ket{S_v T_s^-} & \ket{S_v T_s^+} & \ket{S_v T_s^0} & \ket{T_v^- S_s} &
					\ket{T_v^+ S_s} & \ket{T_v^0 S_s} & \ket{S_v S_s} & \ket{T_v^- T_s^-} &
					\ket{T_v^- T_s^+} & \ket{T_v^- T_s^0} & \ket{T_v^+ T_s^-} & \ket{T_v^+ T_s^+} &
					\ket{T_v^+ T_s^0} & \ket{T_v^0 T_s^-} & \ket{T_v^0 T_s^+} & \ket{T_v^0 T_s^0} \\
					-g_s& 0 & 0 &  0 & 0 & 0 &  0 & 0 & 0 & 0 & 0 & 0 & 0 & 0 & 0 & 0 \\
					  0 &g_s& 0 &  0 & 0 & 0 &  0 & 0 & 0 & 0 & 0 & 0 & 0 & 0 & 0 & 0 \\
					  0 & 0 & 0 &  0 & 0 & 0 &  0 & 0 & 0 & 0 & 0 & 0 & 0 & 0 & 0 & 0 \\
					  0 & 0 & 0 &-g^{(2)}_v& 0 & 0 &  0 & 0 & 0 & 0 & 0 & 0 & 0 & 0 & 0 & 0 \\
					  0 & 0 & 0 &  0 &g^{(2)}_v& 0 &  0 & 0 & 0 & 0 & 0 & 0 & 0 & 0 & 0 & 0 \\
					  0 & 0 & 0 &  0 & 0 & 0 &  0 & 0 & 0 & 0 & 0 & 0 & 0 & 0 & 0 & 0 \\
					  0 & 0 & 0 &  0 & 0 & 0 &  0 & 0 & 0 & 0 & 0 & 0 & 0 & 0 & 0 & 0 \\
					  0 & 0 & 0 &  0 & 0 & 0 &  0 &-g^{(2)}_v-g_s& 0 & 0 & 0 & 0 & 0 & 0 & 0 & 0 \\
					  0 & 0 & 0 &  0 & 0 & 0 &  0 & 0 &-g^{(2)}_v+g_s & 0 & 0 & 0 & 0 & 0 & 0 & 0 \\
					  0 & 0 & 0 &  0 & 0 & 0 &  0 & 0 & 0 &-g^{(2)}_v & 0 & 0 & 0 & 0 & 0 & 0 \\
					  0 & 0 & 0 &  0 & 0 & 0 &  0 & 0 & 0 & 0 &g^{(2)}_v-g_s& 0 & 0 & 0 & 0 & 0 \\
					  0 & 0 & 0 &  0 & 0 & 0 &  0 & 0 & 0 & 0 & 0 &g^{(2)}_v+g_s& 0 & 0 & 0 & 0 \\
					  0 & 0 & 0 &  0 & 0 & 0 &  0 & 0 & 0 & 0 & 0 & 0 &g^{(2)}_v& 0 & 0 & 0 \\
					  0 & 0 & 0 &  0 & 0 & 0 &  0 & 0 & 0 & 0 & 0 & 0 & 0 &-g_s& 0 & 0 \\
					  0 & 0 & 0 &  0 & 0 & 0 &  0 & 0 & 0 & 0 & 0 & 0 & 0 & 0 &g_s& 0 \\
					  0 & 0 & 0 &  0 & 0 & 0 &  0 & 0 & 0 & 0 & 0 & 0 & 0 & 0 & 0 & 0
				\end{pNiceMatrix}
			\end{equation}
		}
	\end{sideways}
	\clearpage
	\begin{sideways}
		\parbox{\textheight}{
			As the valley states $S_v$, $T_v^\mp$, $T_v^0$ differ from each other in reciprocal space, their energy is altered by exchange interaction. This is considered by introducing the exchange splittings $J_1$, $J_2$ usually of the order of a few $\SI{100}{\micro\electronvolt}$:
			\begin{equation}
				H_\mathrm{xc} = \begin{pNiceMatrix}[first-row, nullify-dots]
					\ket{S_v T_s^-} & \ket{S_v T_s^+} & \ket{S_v T_s^0} & \ket{T_v^- S_s} &
					\ket{T_v^+ S_s} & \ket{T_v^0 S_s} & \ket{S_v S_s} & \ket{T_v^- T_s^-} &
					\ket{T_v^- T_s^+} & \ket{T_v^- T_s^0} & \ket{T_v^+ T_s^-} & \ket{T_v^+ T_s^+} &
					\ket{T_v^+ T_s^0} & \ket{T_v^0 T_s^-} & \ket{T_v^0 T_s^+} & \ket{T_v^0 T_s^0} \\
					0 & 0 &  0 & 0 & 0 & 0 &  0 & 0 & 0 & 0 & 0 & 0 & 0 & 0 & 0 & 0 \\
					0 & 0 &  0 & 0 & 0 & 0 &  0 & 0 & 0 & 0 & 0 & 0 & 0 & 0 & 0 & 0 \\
					0 & 0 &  0 & 0 & 0 & 0 &  0 & 0 & 0 & 0 & 0 & 0 & 0 & 0 & 0 & 0 \\
					0 & 0 &  0 &J_1& 0 & 0 &  0 & 0 & 0 & 0 & 0 & 0 & 0 & 0 & 0 & 0 \\
					0 & 0 &  0 & 0 &J_1& 0 &  0 & 0 & 0 & 0 & 0 & 0 & 0 & 0 & 0 & 0 \\
					0 & 0 &  0 & 0 & 0 &J_1+J_2&  0 & 0 & 0 & 0 & 0 & 0 & 0 & 0 & 0 & 0 \\
					0 & 0 &  0 & 0 & 0 & 0 & 0 & 0 & 0 & 0 & 0 & 0 & 0 & 0 & 0 & 0 \\
					0 & 0 &  0 & 0 & 0 & 0 & 0 &J_1& 0 & 0 & 0 & 0 & 0 & 0 & 0 & 0 \\
					0 & 0 &  0 & 0 & 0 & 0 & 0 & 0 &J_1& 0 & 0 & 0 & 0 & 0 & 0 & 0 \\
					0 & 0 &  0 & 0 & 0 & 0 & 0 & 0 & 0 &J_1& 0 & 0 & 0 & 0 & 0 & 0 \\
					0 & 0 &  0 & 0 & 0 & 0 & 0 & 0 & 0 & 0 &J_1& 0 & 0 & 0 & 0 & 0 \\
					0 & 0 &  0 & 0 & 0 & 0 & 0 & 0 & 0 & 0 & 0 &J_1& 0 & 0 & 0 & 0 \\
					0 & 0 &  0 & 0 & 0 & 0 & 0 & 0 & 0 & 0 & 0 & 0 &J_1& 0 & 0 & 0 \\
					0 & 0 &  0 & 0 & 0 & 0 & 0 & 0 & 0 & 0 & 0 & 0 & 0 &J_1+J_2& 0 & 0 \\
					0 & 0 &  0 & 0 & 0 & 0 & 0 & 0 & 0 & 0 & 0 & 0 & 0 & 0 &J_1+J_2& 0 \\
					0 & 0 &  0 & 0 & 0 & 0 & 0 & 0 & 0 & 0 & 0 & 0 & 0 & 0 & 0 & J_1+J_2
				\end{pNiceMatrix}
			\end{equation}
			In order to construct a state that is antisymmetric under exchange, one has to involve an excited orbital state for symmetric valley-spin components. By using the single-particle orbital splitting $\Orb$ we get
			\begin{equation}
				H_\mathrm{Orb} = \Orb\begin{pNiceMatrix}[first-row, nullify-dots]
					\ket{S_v T_s^-} & \ket{S_v T_s^+} & \ket{S_v T_s^0} & \ket{T_v^- S_s} &
					\ket{T_v^+ S_s} & \ket{T_v^0 S_s} & \ket{S_v S_s} & \ket{T_v^- T_s^-} &
					\ket{T_v^- T_s^+} & \ket{T_v^- T_s^0} & \ket{T_v^+ T_s^-} & \ket{T_v^+ T_s^+} &
					\ket{T_v^+ T_s^0} & \ket{T_v^0 T_s^-} & \ket{T_v^0 T_s^+} & \ket{T_v^0 T_s^0} \\
					0 & 0 & 0 & 0 & 0 & 0 & 0 & 0 & 0 & 0 & 0 & 0 & 0 & 0 & 0 & 0 \\
					0 & 0 & 0 & 0 & 0 & 0 & 0 & 0 & 0 & 0 & 0 & 0 & 0 & 0 & 0 & 0 \\
					0 & 0 & 0 & 0 & 0 & 0 & 0 & 0 & 0 & 0 & 0 & 0 & 0 & 0 & 0 & 0 \\
					0 & 0 & 0 & 0 & 0 & 0 & 0 & 0 & 0 & 0 & 0 & 0 & 0 & 0 & 0 & 0 \\
					0 & 0 & 0 & 0 & 0 & 0 & 0 & 0 & 0 & 0 & 0 & 0 & 0 & 0 & 0 & 0 \\
					0 & 0 & 0 & 0 & 0 & 0 & 0 & 0 & 0 & 0 & 0 & 0 & 0 & 0 & 0 & 0 \\
					0 & 0 & 0 & 0 & 0 & 0 & 1 & 0 & 0 & 0 & 0 & 0 & 0 & 0 & 0 & 0 \\
					0 & 0 & 0 & 0 & 0 & 0 & 0 & 1 & 0 & 0 & 0 & 0 & 0 & 0 & 0 & 0 \\
					0 & 0 & 0 & 0 & 0 & 0 & 0 & 0 & 1 & 0 & 0 & 0 & 0 & 0 & 0 & 0 \\
					0 & 0 & 0 & 0 & 0 & 0 & 0 & 0 & 0 & 1 & 0 & 0 & 0 & 0 & 0 & 0 \\
					0 & 0 & 0 & 0 & 0 & 0 & 0 & 0 & 0 & 0 & 1 & 0 & 0 & 0 & 0 & 0 \\
					0 & 0 & 0 & 0 & 0 & 0 & 0 & 0 & 0 & 0 & 0 & 1 & 0 & 0 & 0 & 0 \\
					0 & 0 & 0 & 0 & 0 & 0 & 0 & 0 & 0 & 0 & 0 & 0 & 1 & 0 & 0 & 0 \\
					0 & 0 & 0 & 0 & 0 & 0 & 0 & 0 & 0 & 0 & 0 & 0 & 0 & 1 & 0 & 0 \\
					0 & 0 & 0 & 0 & 0 & 0 & 0 & 0 & 0 & 0 & 0 & 0 & 0 & 0 & 1 & 0 \\
					0 & 0 & 0 & 0 & 0 & 0 & 0 & 0 & 0 & 0 & 0 & 0 & 0 & 0 & 0 & 1
				\end{pNiceMatrix}
			\end{equation}
		}
	\end{sideways}
	\clearpage
	The two-particle hamiltonian is then given by $H^{(2)} = H_\mathrm{SO}+H_\mathrm{Z}+H_\mathrm{xc}+H_\mathrm{Orb}$. With $\Orb \sim \SI{1}{\milli\electronvolt} \gg J_1,J_2 \sim \SI{100}{\micro\electronvolt}, \DSO \approx \SI{80}{\micro\electronvolt}$,  the two particles are assumed to be in the orbital ground state. Hence, we can reduce our considerations to the part of the hamiltonian that contains only the antisymmetric valley-spin states. The low-energy excitation spectrum for two particles is then described by
	\begin{equation}
		\label{eq:hamtwo}
		H^{(2)}_\mathrm{ES} = \begin{pNiceMatrix}[first-row, nullify-dots]
			\ket{S_v T_s^-} & \ket{S_v T_s^+} & \ket{S_v T_s^0} & \ket{T_v^- S_s} & \ket{T_v^+ S_s} & \ket{T_v^0 S_s}  \\
			-g_s\muB\Bper&  0 &  0  & 0 & 0 &  0 \\
			0 & g_s\muB\Bper&  0  & 0 & 0 &  0 \\
			0 &  0 & 0 & 0 & 0 &-\DSO\\
			0 &  0 & 0  &J_1-g^{(2)}_v\muB\Bper & 0 &  0  \\
			0 &  0 & 0  & 0 &J_1+g_v^{(2)}\muB\Bper &  0  \\
			0 &  0 &-\DSO& 0 & 0 & J_1+J_2 \\
		\end{pNiceMatrix}
	\end{equation}

	The eigen states and energies of Eq.~\ref{eq:hamone} are compiled in Tab.~\ref{tab:onespectrum}, the eigen states and energies of Eq.~\ref{eq:hamtwo} are compiled in Tab.~\ref{tab:twospectrum}. Their evolution in magnetic field are visualized in Fig.~\ref{fig:oldnewspectrum}.

	\begin{table*}[h!]
		\begin{minipage}{0.48\textwidth}
			\caption{\label{tab:onespectrum}Excited state spectrum of the one-carrier charge state.}
			\begin{ruledtabular}
				\begin{tabular}{cc}
					eigen states & eigen energies (w.r.t. $E^{(1)}_0$)\\
					\colrule
					$\ket{K^-\downarrow}$ & $-\frac{1}{2} \DSO-\frac{1}{2} (g_v^{(1)} + g_s)\muB \Bper$ \\
					$\ket{K^+\uparrow}$ & $-\frac{1}{2} \DSO + \frac{1}{2} (g_v^{(1)} + g_s)\muB \Bper$ \\
					\colrule \\
					$\ket{K^-\uparrow}$ & $\frac{1}{2} \DSO-\frac{1}{2} (g_v^{(1)} - g_s)\muB \Bper$ \\
					$\ket{K^+\downarrow}$ & $\frac{1}{2} \DSO+\frac{1}{2} (g_v^{(1)} - g_s)\muB \Bper$ \\
				\end{tabular}
			\end{ruledtabular}
		\end{minipage}\hfill 
		\begin{minipage}{0.48\textwidth}
			\caption{\label{tab:twospectrum}Excited state spectrum of the two-carrier charge state.}
			\begin{ruledtabular}
				\begin{tabular}{cc}
					eigen states & eigen energies (w.r.t. $E^{(2)}_0$) \\
					\colrule
					$\cos(\frac{\theta}{2}) \ket{S_v T_s^0} +
					\sin(\frac{\theta}{2}) \ket{T_v^0 S_s}$ & \\
					$=: \ket{S_v T_s^0} \oplus \ket{T_v^0 S_s}$ & $-\DSO'$ \\
					\colrule
					$\ket{S_v T_s^-}$ & $-g_s\muB\Bper$ \\
					$\ket{S_v T_s^0}$ & $0$ \\
					$\ket{S_v T_s^+}$ & $g_s\muB\Bper$ \\
					\colrule
					$\ket{S_s T_v^-}$ & $J_1-g_v^{(2)}\muB\Bper$ \\
					$\ket{S_s T_v^+}$ & $J_1+g_v^{(2)}\muB\Bper$ \\
					\colrule
					$\ket{S_s T_v^0}$ & $J_1+J_2$ \\
					\colrule
					$-\sin(\frac{\theta}{2}) \ket{S_v T_s^0} +
					\cos(\frac{\theta}{2}) \ket{T_v^0 S_s}$ & \\
					$=: \ket{S_v T_s^0} \boxplus \ket{T_v^0 S_s}$ &
					$J_1 + J_2 + \DSO'$ \\
				\end{tabular}
		\end{ruledtabular}
		\end{minipage}
	\end{table*}

	\begin{figure}[b!]
		\includegraphics{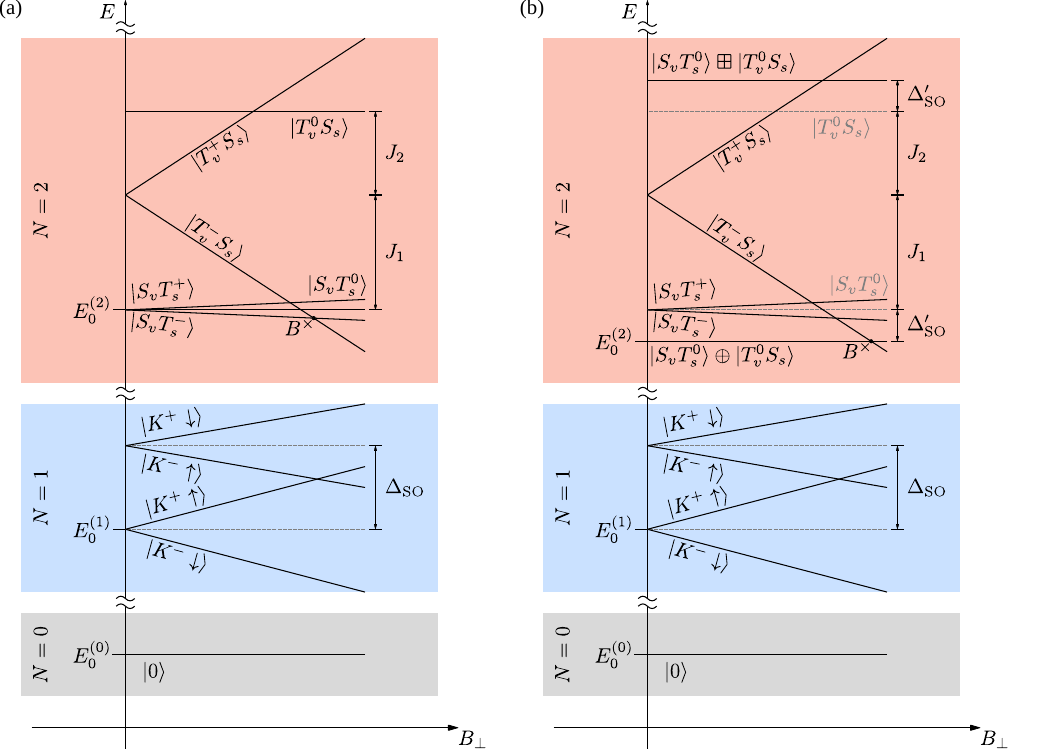}
		\caption{\label{fig:oldnewspectrum} Energy spectrum without (a) and with (b) considering the effect of Kane-Mele type SOI for the two-carrier charge state $N=2$.}
	\end{figure}
	The splitting induced by the Kane-Mele gap $\DSO$ is given by
	\begin{equation}
		\DSO' = \frac{J_1+J_2}{2} \left(\sqrt{\left(\frac{2\DSO}{J_1 + J_2}\right)^2 + 1} - 1 \right) > 0
	\end{equation}
	The angle $\theta \in \interval{0}{\frac{\pi}{2}}$ characteristic for the superposition that forms the new ground state is given by the relation
	\begin{equation}
		\label{eq:theta}
		\tan(\theta) = \frac{2\DSO}{J_1 + J _2}
	\end{equation}
	From this one can see that the strength of mixing between states $\ket{S_v T_s^0}$ and $\ket{T_v^0 S_s}$ depends on how large the Kane-Mele spin--orbit gap $\DSO$ is compared to the exchange splittings $J_1$, $J_2$. For $\DSO \ll J_1 + J_2$ one obtains $\ket{S_v T_s^0}$ as the ground state, whereas for $\DSO \gg J_1 + J_2$ the ground state is given by the superposition $\frac{1}{\sqrt{2}}(\ket{S_v T_s^0} + \ket{T_v^0 S_s})$.

	Applying an out-of-plane magnetic field $\Bper$ does not cause a change in ground state for the $N=0$ and $N=1$ charge carrier states. For the $N=2$ charge carrier state there is a ground state crossing happening at field $B^\times$. For $\Bper > B^\times$ the new ground state is given by $\ket{T_v^- S_s}$.

	\clearpage

\end{onecolumngrid} \clearpage
	\begin{onecolumngrid}
	\section{Finite bias spectroscopy}
	\label{app:spectroscopy}

		\begin{figure}[h!]
			\includegraphics{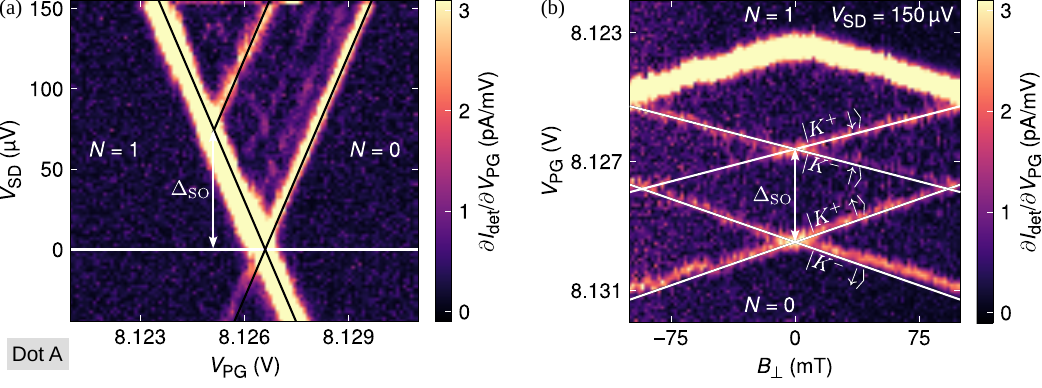}
			\caption{\label{fig:onehole} (a) Coulomb diamond measurement at the $0\to 1$-transition. The transconductance peaks show the ground state and an excited state separated in energy $\DSO$ from the ground state. The lever arm for the first carrier is extracted from this measurement. (b) Finite bias measurement in out-of-plane magnetic field. The ground state splits into two states, the Kramers pair $\ket{K^-\downarrow}$, $\ket{K^+\uparrow}$. The excited state splits into the Kramers pair $\ket{K^-\uparrow}$, $\ket{K^+\downarrow}$. From the slopes we extract the $g$-factors for the valley- $g_v^{(1)}$ and spin-Zeeman effect $g_s^{(1)}$ for the first carrier. All extracted parameters are compiled in Tab.~\ref{tab:oneparams}.}
		\end{figure}

		\begin{table*}[h!]
			\begin{minipage}{0.48\textwidth}
				\caption{\label{tab:oneparams} Parameters from $0 \to 1$ transition}
				\begin{ruledtabular}
					\begin{tabular}{ll}
						lever arm & $\apg^{(1)} = 0.025$ \\
						\colrule
						SO splitting & $\DSO = \SI{73}{\micro\electronvolt}$ \\
						\colrule
						valley $g$-factor & $g_v^{(1)} = 13.6$ \\
						spin $g$-factor & $g_s^{(1)} = 2.0$ \\
					\end{tabular}
				\end{ruledtabular}
			\end{minipage}\hfill 
			\begin{minipage}{0.48\textwidth}
				\caption{\label{tab:twoparams} Parameters from $1 \to 2$ transition}
				\begin{ruledtabular}
					\begin{tabular}{ll}
						leverarm & $\apg^{(2)} = 0.025$ \\
						\colrule
						SO induced splitting & $\DSO' = \SI{45}{\micro\electronvolt}$ \\
						exchange splitting & $J_1 = \SI{160}{\micro\electronvolt}$ \\
						\colrule
						valley $g$-factors first carrier & $g_v^{(1)} = 15.5$ \\
						valley $g$-factor second carrier & $g_v^{(2)} = 16.0$ \\
						spin $g$-factor & $g_s^{(2)} = 2.0$ \\
					\end{tabular}
				\end{ruledtabular}
			\end{minipage}
		\end{table*}

		\clearpage
		
		\begin{figure}[h!]
			\includegraphics{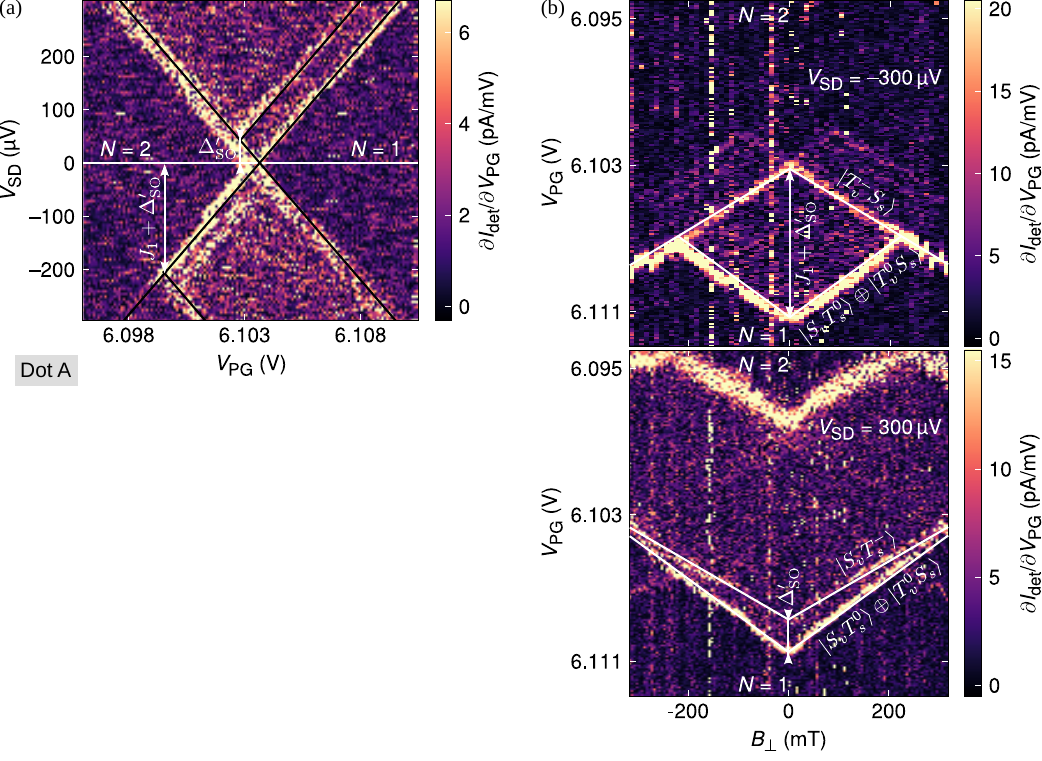}
			\caption{\label{fig:twohole} (a) Coulomb diamond measurement at the $1\to 2$-transition. The transconductance peaks show the ground state and two excited states: one for positive bias separated by theenergy $\DSO'$ from the ground state, one for negative bias separated by the energy $\DSO'+J_1$ from the ground state. The lever arm for the first carrier is extracted from this measurement. (b) Finite bias measurement in out-of-plane magnetic field. The ground state does not split. The excited state with energy separation of $\DSO'$ only visible for positive bias moves down in energy with a $g$-factor of $2.0$, indicating that it is the state$\ket{S_v T_s^-}$. The excited state with an energy separation of $\DSO'+J_1$ only visible for negative bias moves down in energy with a $g$-factor of $16.0$, indicating that it is the state$\ket{S_v T_s^-}$. All extracted parameters are compiled in Tab.~\ref{tab:twoparams}.}
		\end{figure}
		
\end{onecolumngrid} \clearpage
	\def\DeSO{\ensuremath{\Delta_\mathrm{eSO}}}

\begin{onecolumngrid}
	\section{\label{app:electron} Device B: Electron QD}

	\begin{figure*}[h!]
		\includegraphics{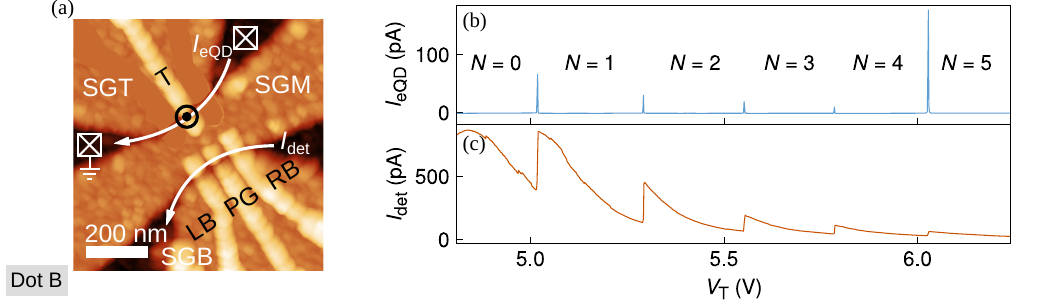}
		\caption{\label{fig:electronstates}
			(a) The electron QD is electrostatically formed under finger gate $T$. Carriers are confined by pn-junctions at a displacement field of $D_e = -\SI{0.21}{\volt\per\nano\meter}$ (contrary to $D = -\SI{0.46}{\volt\per\nano\meter}$ for the hole QD in the main text). (b) Coulomb blockade peaks and (c) charge detection signal for the first five electrons.
		}
	\end{figure*}
    
	\begin{figure*}[h!]
		\includegraphics{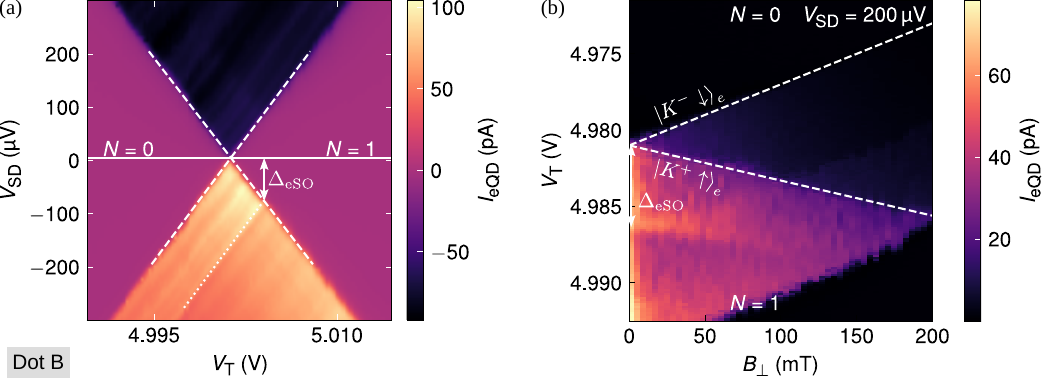}
		\caption{\label{fig:1stelectron}
		(a) Coulomb blockade diamond data of the first electron at zero magnetic field. The extracted leverarm is $\alpha_\mathrm{T} = 0.0147$, the Kane-Mele spin-orbit splitting is $\DeSO = \SI{80}{\micro\electronvolt}$, agreeing with the observed gap in the hole-type QD. (b) Excite state spectroscopy data at constant bias in magnetic field. The extracted $g$-factor of the valley Zeeman effect is $g_v = 14$.
		}
	\end{figure*}

	\begin{figure*}[h!]
		\includegraphics{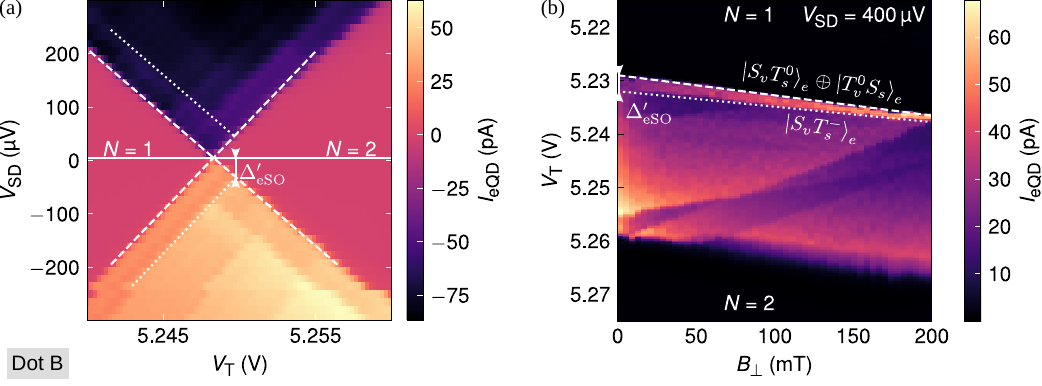}
		\caption{\label{fig:2ndelectron}
			(a) Coulomb blockade diamond data of the first electron at zero magnetic field. The extracted leverarm is $\alpha_\mathrm{T} = 0.0131$, the gap between the ground state and first excited state is $\DeSO' = \SI{40}{\micro\electronvolt}$. (b) Excite state spectroscopy data at constant bias in magnetic field. The first excited state (dotted line) approaches the ground state (dashed line) with a $g$-factor of $2$, confirming the proposed spectrum in Fig.~\ref{fig:spectrum}.
		}
	\end{figure*}
\end{onecolumngrid} \clearpage
	\begin{onecolumngrid}

		\section{Thermodynamic Model}
		\label{app:entropy}
		\subsection{Grand canonical ensemble}
		QD is coupled to a reservoir, allowing for carrier and energy to fluctuate. If the coupling is dominated by thermal broadening, i.e. $\kB T \gg \GammaR$, the thermodynamics of the system is in general described in the framework of the grand canonical ensemble. The grand partition function is given by
		\begin{align}
			\Zfun(\mu,T)
			&= \sum_{N=0}^{\infty} \sum_{i=0}^{\infty} \euler^{-\frac{E^{(N)}_i - N \mu}{\kB T}} \\
			&= \sum_{N=0}^{\infty} \euler^{-\frac{E^{(N)}_0 - N \mu}{\kB T}} \sum_{i=0}^{\infty} \euler^{-\frac{\Delta^{(N)}_i}{\kB T}}
			\label{eq:Zfun}
		\end{align}
		where we introduced the decomposition
		\begin{equation}
			E^{(N)}_i = E^{(N)}_0 + \Delta^{(N)}_i
		\end{equation}
		where $E^{(N)}_{i}$ is the energy of the $i$th excited state of the $N$-charge carrier state, $E^{(N)}_{0}$ the ground state energy of the $N$-charge carrier state, and $\Delta^{(N)}_{i}$ the energy difference between the respective excited state and the ground state.
		All macroscopic state quantities such as $\Nbar(\mu,T)$ and $S(\mu,T)$ can be calculated from $\Zfun(\mu,T)$, as well as the occupation probability $p^{(N)}_i(\mu,T)$ of the microstate labeled $(N,i)$. In terms of $\Zfun(\mu,T)$ one finds
		\begin{align}
			p^{(N)}_i(\mu,T)
			&= -\kB T \diffp{\ln\Zfun(\mu,T)}{E^{(N)}_i} \\
			\Nbar(\mu,T)
			&= \kB T \diffp{\ln\Zfun(\mu,T)}{\mu} \\
			S(\mu,T)
			&=  \kB \ln\Zfun(\mu,T) + \kB T \diffp{\ln\Zfun(\mu,T)}{T} \text{ .}
		\end{align}
		and in terms of microstates/microstate probabilities one has
		\begin{align}
			p^{(N)}_i(\mu,T)
			&= \frac{\euler^{-\frac{E^{(N)}_0 - N \mu}{\kB T}}}{\Zfun(\mu,T)} \euler^{-\frac{\Delta^{(N)}_i}{\kB T}} \label{eq:microprob}\\
			\Nbar(\mu,T)
			&= \sum_{N=0}^{\infty} N \sum_{i=0}^\infty p^{(N)}_i(\mu,T) \\
			S(\mu,T)
			&= -\kB \sum_{N=0}^{\infty} \sum_{i=0}^\infty p^{(N)}_i \ln(p^{(N)}_i)
			\label{eq:microentropy}
		\end{align}

		\subsection{Entropy at zero magnetic field}
		For the charge transition $N-1 \leftrightarrow N$ contributions of other charge states can be neglected if $\kB T \ll E^{(N-1)}_0-E^{(N-2)}_0\text{, } E^{(N+1)}_0-E^{(N)}_0$. The grand partition function becomes
		\begin{equation}
			\Zfun^{(N-1\leftrightarrow N)}(\mu,T)
			= \euler^{-\frac{E^{(N-1)}_0 - (N-1) \mu}{\kB T}} \sum_{i=0}^{\infty} \euler^{-\frac{\Delta^{(N-1)}_i}{\kB T}}
			+ \euler^{-\frac{E^{(N)}_0 - N \mu}{\kB T}} \sum_{i=0}^{\infty} \euler^{-\frac{\Delta^{(N)}_i}{\kB T}}
			\label{eq:Ztrans}
		\end{equation}
		If additionally $\kB T \ll \Delta^{(N)}_i\text{, } \Delta^{(N-1)}_i$ one can also neglect excited states of the respective charge state in the expression for the grand partition function. Under these considerations, the grand partition function, the microstate probabilities, and the entropy are for the $0 \to 1$-transition given by
		
		\begin{align}
			\Zfun^{(0\leftrightarrow 1)}
			&= \euler^{-\frac{E^{(0)}_0}{\kB T}}
			+ \euler^{-\frac{E^{(1)}_0 - \mu}{\kB T}} \cdot 2 \\
			p^{(0)}_0
			&= \frac{1}{1 + \euler^{-\frac{\mu^{(1)}_0 - \mu}{\kB T} + \ln(2)}} \\
			p^{(1)}_{\ket{K^-\downarrow}}
			&= p^{(1)}_{\ket{K^+\uparrow}}
			= \frac{1}{2} \frac{1}{1 + \euler^{\frac{\mu^{(1)}_0 - \mu}{\kB T} - \ln(2)}} \\
			S^{(0\leftrightarrow 1)}
			&= -\kB \left[p^{(0)}_0 \ln(p^{(0)}_0) + p^{(1)}_{\ket{K^-\downarrow}} \ln(p^{(1)}_{\ket{K^-\downarrow}}) + p^{(1)}_{\ket{K^+\uparrow}} \ln(p^{(1)}_{\ket{K^+\uparrow}}) \right] \label{eq:S01}
		\end{align}
		
		and for the $1 \to 2$-transition
		
		\begin{align}
			\Zfun^{(1\leftrightarrow 2)}
			&= \euler^{-\frac{E^{(1)}_0 - \mu}{\kB T}} \cdot 2
			+ \euler^{-\frac{E^{(2)}_0 - 2\mu}{\kB T}} \\
			p^{(1)}_{\ket{K^-\downarrow}}
			= p^{(1)}_{\ket{K^+\uparrow}}
			&= \frac{1}{2} \frac{1}{1 + \euler^{-\frac{\mu^{(2)}_0 - \mu}{\kB T} - \ln(2)}} \\
			p^{(2)}_{\ket{S_v T_s^0} \oplus \ket{T_v^0 S_s}}
			&= \frac{1}{1 + \euler^{\frac{\mu^{(2)}_0 - \mu}{\kB T} 
					+ \ln(2)}} \\
			S^{(1\leftrightarrow 2)}
			&= -\kB \left[p^{(1)}_{\ket{K^-\downarrow}} \ln(p^{(1)}_{\ket{K^-\downarrow}}) + p^{(1)}_{\ket{K^+\uparrow}} \ln(p^{(1)}_{\ket{K^+\uparrow}}) + p^{(2)}_{\ket{S_v T_s^0} \oplus \ket{T_v^0 S_s}} \ln(p^{(2)}_{\ket{S_v T_s^0} \oplus \ket{T_v^0 S_s}}) \right] \label{eq:S12}
		\end{align}
		
		Note that here the kets merely serve as labels. Only the ground state degeneracies of the respective charge states and the temperature determine the line shape of these functions.

		\subsection{Entropy at fixed particle number $N$}
		By putting the chemical potential $\mu$ far away from any transition, the particle number $N$ is fixed. The QD is in the $N$ carrier state with absolute certainty. A description in the canonical ensemble framework is appropriate. The partition function is then given by
		\begin{equation}
			Z^{(N)}(T)
			= \sum_{j=0}^{\infty} \euler^{-\frac{\Delta^{(N)}_j}{\kB T}}
		\end{equation}
		The probabilities of the microstates being realized is
		\begin{equation}
			p^{(N)}_i(T)
			= \frac{\euler^{-\frac{\Delta^{(N)}_i}{\kB T}}}{\sum\limits_{j=0}^{\infty} \euler^{-\frac{\Delta^{(N)}_j}{\kB T}}}
			\label{eq:microprob} 
		\end{equation}

		\subsection{Microstate probabilities of the BLG QD}

		For the $N=0$ carrier state we find
		\begin{align}
			Z^{(0)} &= 1 \\
			p^{(0)}_0 &= 1 \\
			S_0 &= 0
		\end{align}
		With Tab.~\ref{tab:onespectrum} we find for the $N=1$ carrier state
		\begin{align}
			Z^{(1)} &= \euler^{\frac{(g_v^{(1)} + g_s)\muB \Bper}{2\kB T}} +\euler^{\frac{(g_v^{(1)} + g_s)\muB \Bper}{2\kB T}} \\
			p^{(1)}_{\ket{K^-\downarrow}} &= \frac{\euler^{\frac{(g_v^{(1)} + g_s)\muB \Bper}{2\kB T}}}{Z^{(1)}} \\
			p^{(1)}_{\ket{K^+\uparrow}} &= \frac{\euler^{-\frac{(g_v^{(1)} + g_s)\muB \Bper}{2\kB T}}}{Z^{(1)}} \\
			S_1 &= -\kB \left[p^{(1)}_{\ket{K^-\downarrow}} \ln(p^{(1)}_{\ket{K^-\downarrow}}) + p^{(1)}_{\ket{K^+\uparrow}} \ln(p^{(1)}_{\ket{K^+\uparrow}}) \right]
		\end{align}
		where we included only the states lowest in energy as $\DSO > 4\kB T$ in the experiments.
		For the $N=2$ carrier state we find with Tab.~\ref{tab:twospectrum}:
		\begin{align}
			Z^{(2)} &= 1 +\euler^{-\frac{\DSO' + J_1 - g_v^{(2)}\muB \Bper}{\kB T}} \\
			p^{(2)}_{\ket{S_v T_s^0} \oplus \ket{T_v^0 S_s}} &= \frac{1}{Z^{(2)}} \\
			p^{(2)}_{\ket{T_v^- S_s}} &= \frac{\euler^{-\frac{\DSO' + J_1 - g_v^{(1)}\muB \Bper}{\kB T}}}{Z^{(2)}} \\
			S_2 &= -\kB \left[p^{(2)}_{\ket{S_v T_s^0} \oplus \ket{T_v^0 S_s}} \ln(p^{(2)}_{\ket{S_v T_s^0} \oplus \ket{T_v^0 S_s}}) + p^{(2)}_{\ket{T_v^- S_s}} \ln(p^{(2)}_{\ket{T_v^- S_s}}) \right]
		\end{align}
		where we included only the states lowest in energy as $\DSO' > 4 \kB T$ and states causing a ground state crossing at finite $\Bper$. With these expressions we define
		\begin{align}
			\dl.Delta.{S_{0\to 1}^\mathrm{calc}} &= S_1 - S_0 \label{eq:dS01} \\
			\dl.Delta.{S_{1\to 2}^\mathrm{calc}} &= S_2 - S_1 \label{eq:dS12}
		\end{align}
		
\end{onecolumngrid}

\end{document}